\documentclass[iop]{emulateapj}
\usepackage{apjfonts}
\usepackage{graphicx}
\usepackage{amsmath}
\usepackage{epstopdf}

\slugcomment{Accepted for publication in ApJS}
\begin{document}

\title{4.5 years multi-wavelength observations of Mrk 421 during the ARGO-YBJ and Fermi common operation time}

%\author{\centerline {{\it The ARGO-YBJ Collaboration: }}}
%\author{\centerline {{\it The ARGO-YBJ Collaboration: }}}
\author{B.~Bartoli\altaffilmark{1,2},
 P.~Bernardini\altaffilmark{3,4},
 X.J.~Bi\altaffilmark{5},
 Z.~Cao\altaffilmark{5},
 S.~Catalanotti\altaffilmark{1,2},
 S.Z.~Chen\altaffilmark{5,*},%\footnotetext[0]{Corresponding author: Songzhan Chen, chensz@ihep.ac.cn}
 T.L.~Chen\altaffilmark{6},
 S.W.~Cui\altaffilmark{7},
 B.Z.~Dai\altaffilmark{8},
 A.~D'Amone\altaffilmark{3,4},
 Danzengluobu\altaffilmark{6},
 I.~De Mitri\altaffilmark{3,4},
 B.~D'Ettorre Piazzoli\altaffilmark{1,2},
 T.~Di Girolamo\altaffilmark{1,2},
 G.~Di Sciascio\altaffilmark{9},
 C.F.~Feng\altaffilmark{10},
 Zhaoyang Feng\altaffilmark{5},
 Zhenyong Feng\altaffilmark{11},
 Q.B.~Gou\altaffilmark{5},
 Y.Q.~Guo\altaffilmark{5},
 H.H.~He\altaffilmark{5},
 Haibing Hu\altaffilmark{6},
 Hongbo Hu\altaffilmark{5},
 M.~Iacovacci\altaffilmark{1,2},
 R.~Iuppa\altaffilmark{9,12},
 H.Y.~Jia\altaffilmark{11},
 Labaciren\altaffilmark{6},
 H.J.~Li\altaffilmark{6},
 C.~Liu\altaffilmark{5},
 J.~Liu\altaffilmark{8},
 M.Y.~Liu\altaffilmark{6},
 H.~Lu\altaffilmark{5},
 L.L.~Ma\altaffilmark{5},
 X.H.~Ma\altaffilmark{5},
 G.~Mancarella\altaffilmark{3,4},
 S.M.~Mari\altaffilmark{13,14},
 G.~Marsella\altaffilmark{3,4},
 S.~Mastroianni\altaffilmark{2},
 P.~Montini\altaffilmark{9},
 C.C.~Ning\altaffilmark{6},
 L.~Perrone\altaffilmark{3,4},
 P.~Pistilli\altaffilmark{13,14},
 P.~Salvini\altaffilmark{15},
 R.~Santonico\altaffilmark{9,12},
 P.R.~Shen\altaffilmark{5},
 X.D.~Sheng\altaffilmark{5},
 F.~Shi\altaffilmark{5},
 A.~Surdo\altaffilmark{4},
 Y.H.~Tan\altaffilmark{5},
 P.~Vallania\altaffilmark{16,17},
 S.~Vernetto\altaffilmark{16,17},
 C.~Vigorito\altaffilmark{17,18},
 H.~Wang\altaffilmark{5},
 C.Y.~Wu\altaffilmark{5},
 H.R.~Wu\altaffilmark{5},
 L.~Xue\altaffilmark{10},
 Q.Y.~Yang\altaffilmark{8},
 X.C.~Yang\altaffilmark{8},
 Z.G.~Yao\altaffilmark{5},
 A.F.~Yuan\altaffilmark{6},
 M.~Zha\altaffilmark{5},
 H.M.~Zhang\altaffilmark{5},
 L.~Zhang\altaffilmark{8},
 X.Y.~Zhang\altaffilmark{10},
 Y.~Zhang\altaffilmark{5},
 J.~Zhao\altaffilmark{5},
 Zhaxiciren\altaffilmark{6},
 Zhaxisangzhu\altaffilmark{6},
 X.X.~Zhou\altaffilmark{11},
 F.R.~Zhu\altaffilmark{11}, and
 Q.Q.~Zhu\altaffilmark{5}\\ (The ARGO-YBJ Collaboration)}


\altaffiltext{*}{Corresponding author: Songzhan Chen, chensz@ihep.ac.cn}
 \affil{  \altaffilmark{1}Dipartimento di Fisica dell'Universit\`a di Napoli
                  ``Federico II'', Complesso Universitario di Monte
                  Sant'Angelo, via Cinthia, 80126 Napoli, Italy.}
  \affil{\altaffilmark{2}Istituto Nazionale di Fisica Nucleare, Sezione di
                  Napoli, Complesso Universitario di Monte
                  Sant'Angelo, via Cinthia, 80126 Napoli, Italy.}
  \affil{\altaffilmark{3}Dipartimento Matematica e Fisica "Ennio De Giorgi",
                  Universit\`a del Salento,
                  via per Arnesano, 73100 Lecce, Italy.}
 \affil{ \altaffilmark{4}Istituto Nazionale di Fisica Nucleare, Sezione di
                  Lecce, via per Arnesano, 73100 Lecce, Italy.}
 \affil{ \altaffilmark{5}Key Laboratory of Particle Astrophysics, Institute
                  of High Energy Physics, Chinese Academy of Sciences,
                  P.O. Box 918, 100049 Beijing, P.R. China.}
 \affil{ \altaffilmark{6}Tibet University, 850000 Lhasa, Xizang, P.R. China.}
  \affil{\altaffilmark{7}Hebei Normal University,  050024, Shijiazhuang
                   Hebei, P.R. China.}
 \affil{ \altaffilmark{8}Yunnan University, 2 North Cuihu Rd., 650091 Kunming,
                   Yunnan, P.R. China.}
 \affil{ \altaffilmark{9}Istituto Nazionale di Fisica Nucleare, Sezione di
                  Roma Tor Vergata, via della Ricerca Scientifica 1,
                  00133 Roma, Italy.}
  \affil{\altaffilmark{10}Shandong University, 250100 Jinan, Shandong, P.R. China.}
  \affil{\altaffilmark{11}Southwest Jiaotong University, 610031 Chengdu,
                   Sichuan, P.R. China.}
 \affil{ \altaffilmark{12}Dipartimento di Fisica dell'Universit\`a di Roma
                  ``Tor Vergata'', via della Ricerca Scientifica 1,
                  00133 Roma, Italy.}
 \affil{ \altaffilmark{13}Dipartimento di Fisica dell'Universit\`a ``Roma Tre'',
                   via della Vasca Navale 84, 00146 Roma, Italy.}
 \affil{ \altaffilmark{14}Istituto Nazionale di Fisica Nucleare, Sezione di
                  Roma Tre, via della Vasca Navale 84, 00146 Roma, Italy.}
 \affil{ \altaffilmark{15}Istituto Nazionale di Fisica Nucleare, Sezione di Pavia,
                   via Bassi 6, 27100 Pavia, Italy.}
  \affil{\altaffilmark{16}Osservatorio Astrofisico di Torino dell'Istituto Nazionale
                   di Astrofisica, via P. Giuria 1, 10125 Torino, Italy.}
 \affil{ \altaffilmark{17}Istituto Nazionale di Fisica Nucleare,
                   Sezione di Torino, via P. Giuria 1, 10125 Torino, Italy.}
 \affil{ \altaffilmark{18}Dipartimento di Fisica dell'Universit\`a di
                   Torino, via P. Giuria 1, 10125 Torino, Italy.}

\begin{abstract}
We report on the extensive multi-wavelength observations of the blazar Markarian 421 (Mrk 421) covering radio to $\gamma$-rays, during the 4.5 year period of ARGO-YBJ and Fermi common operation time, from August 2008 to February 2013. These long-term observations extending over an energy range of 18 orders of magnitude provide a unique chance to study the variable emission of Mrk 421. In particular, thanks to the ARGO-YBJ and $Fermi$ data, the whole energy range from 100 MeV to 10 TeV is covered without any gap. In the observation period, Mrk 421 showed both low and high activity states at all wavebands.
The correlations among flux variations in different wavebands were analyzed.
The X-ray flux is clearly correlated with the TeV $\gamma$-ray flux, while GeV $\gamma$-rays only show a partial correlation with TeV $\gamma$-rays. Radio and UV fluxes seem to be weakly or not correlated with the X-ray and $\gamma$-ray fluxes.
Seven large flares, including five X-ray flares and two GeV $\gamma$-ray flares with variable durations (3$-$58 days), and one X-ray outburst phase were identified and used to investigate the variation of the spectral energy distribution with respect to a relative quiescent phase.  During the outburst phase and the seven flaring episodes, the peak energy in X-rays is observed to increase from sub-keV  to   few keV. The TeV $\gamma$-ray flux  increases up to 0.9$-$7.2 times the flux of the Crab Nebula. The behavior of GeV $\gamma$-rays is found to vary depending on the flare, a feature that leads us to classify flares into three groups according to the GeV flux variation. Finally, the one-zone synchrotron self-Compton model was adopted to describe the emission spectra. Two out of three groups can be satisfactorily described using injected electrons with a power-law spectral index around 2.2, as expected from relativistic diffuse shock acceleration, whereas the remaining group requires a harder injected spectrum.  The underlying physical mechanisms responsible for different groups may be related to the acceleration process or to the environment properties.
\end{abstract}


\section{Introduction}
Active galactic Nuclei (AGNs), one of the most luminous sources of electromagnetic radiation
in the universe, are galaxies with a strong and variable
non-thermal emission, believed to be the result of accretion of mass onto a supermassive black hole
(with a mass ranging from $\sim$10$^6$ to $\sim$10$^{10}$ M$_{\bigodot}$) lying at the center of the galaxy.
In some cases ($\leq$10\%) AGNs show powerful and highly collimated relativistic
jets shooting out in opposite directions, perpendicular to the accretion disc.
The jets emanate from the vicinity of the black hole ($\sim$0.1 pc) and extend up to $\sim$ 1 Mpc.
They are usually associated to several bright superluminal knots, which appear related to
the episodic ejection of plasmoid blobs
(see for example the case of the active galaxy M87 \citep{chenug07}).
The origin of the AGN jets is one of the open problems in astrophysics.

AGNs viewed at a small angle to the axis of the jet are called blazars. They usually show flat radio spectra, strong variability, optical polarization and $\gamma$-ray emission. Blazars include BL Lac objects, which have a lower luminosity and lack of strong emission lines in the optical band, and Flat-Spectrum Radio Quasars (FSRQ), which show a higher luminosity with strong and broad emission lines. The strongly Doppler-boosted radiation makes blazars
the most extreme class of AGNs, where the boosted emission overwhelms all other emissions
from the source. Therefore, the observation of blazars allows a deep insight into the physical
conditions and emission processes of relativistic jets.

Blazars are the dominant extragalactic source class in $\gamma$-rays,
as revealed by the $Fermi$  Large Area Telescope (LAT)  survey at GeV energies \citep{nolan12}.
Moving to very high energies (VHE, $>$0.1 TeV), the BL Lac objects dominate the extragalactic sky.
Up to now, 60 AGNs have been established as VHE $\gamma$-ray emitters,
including 52 BL Lac objects\footnote{http://tevcat.uchicago.edu/(Version: 3.400, as of March 2015).}.
Although the $\gamma$-ray emission from blazars has been studied for about two decades,
it is still unclear where and how the emission originates.
Observations of the misaligned radio galaxy M87 indicate that
VHE $\gamma$-rays, at least during flaring periods, seem to originate within
the jet collimation region, in the immediate vicinity of the black hole \citep{accia09,abram12}.
The high energy particles responsible for the nonthermal emission are generally believed
to be accelerated in the relativistic shock front, described by the theory of
diffusive acceleration (for a review, see \cite{kirk99}).

The radiation of a blazar is a broadband continuum ranging from radio through X-rays to $\gamma$-rays.
The spectral energy distributions (SEDs) are characterized by two distinct bumps, which are believed
to be dominated by non-thermal emission. The lower energy component, which peaks in the optical through X-ray,
is caused by the synchrotron radiation from relativistic electrons (and positrons) within the jet.
The origin of the high energy $\gamma$-ray component is still debated.
The general view attributes it to inverse Compton scattering of the synchrotron
(synchrotron self-Compton, SSC) or external photons (external Compton, EC) by the same population of relativistic electrons \citep{ghise98,dermer92,sikor94}.
However, the hadronic scenario, which attributes the $\gamma$-ray emission
to proton-initiated cascades and/or to proton-synchrotron emission in a magnetic field-dominated jet \citep{aharo00}, cannot be excluded.

In this panorama, multi-wavelength observations are of fundamental importance. According
to the present measurements, X-rays and  VHE $\gamma$-rays are correlated during the flaring periods
(for reviews, see \cite{wagner08,chen13}).
Recently, a long-term continuous monitoring of Mrk 421 performed by the  Astrophysical Radiation with Ground-based Observatory at YangBaJing (ARGO-YBJ)
experiment and different satellite-borne X-ray detectors \citep{barto11} showed a good
correlation in terms of flux and spectral index.
All the observational features indicate that $\gamma$-rays and X-rays have a common
origin, supporting the leptonic models. The tight correlation is
a challenge for models based on hadronic processes.
According to a recent collective evidence \citep{meyer12}, the SSC mechanism seems to dominate the emission of BL Lac objects, while the EC component becomes important for FSRQ.
The lack of strong emission lines in the radiation of BL Lac objects is also taken as an
evidence for a minor role of ambient photons (e.g., \cite{krawczy04}), favouring the SSC model.
In this sense we can assume that BL Lac objects are less affected by the circumambient background
radiation and can be considered as ideal targets for the study of the physical processes within
the jets. However, even in the framework of the SSC model, the fundamental question of
the origin of the flux and spectral variability, observed on timescales from minutes to years,
is still open.

Mrk 421 (z=0.031), classified as a BL Lac object, is one of the brightest VHE $\gamma$-ray blazars known. It is a very active blazar with major outbursts, composed of many short flares, about once every two years, in both X-rays and $\gamma$-rays \citep{chen13,aielli10,barto11}.
Actually it is considered  an excellent candidate to study  the physical processes within the AGN jets.
During the last decade, several coordinated multi-wavelength campaigns focusing on Mrk 421 were conducted, both in response to strong outbursts or as part of dedicated observation campaigns.
Complex relations between X-rays and VHE $\gamma$-rays spectra were observed in many flares (see our previous review in \citep{barto11}). However, due to the sparse multi-frequency data during long periods of time, no systematic studies on flux variation and SED evolution were achieved, especially in the  $\gamma$-ray band.
In the beginning of 2009, a multi-frequency observational campaign of Mrk 421 was carried
out for 4.5 months with an excellent temporal (except at VHE) and energy coverage from radio to
VHE $\gamma$-rays. During the whole campaign, however, Mrk 421 showed a low activity at all wavebands \citep{abdo11}.

To understand the emission variability and the underlying acceleration and radiation mechanisms in jets, continuous multi-wavelength observations, particularly in X-rays and VHE $\gamma$-rays, are crucial.
A simultaneous SED could
provide  a snapshot of the emitting population of particles and also constrain the model parameters at a given time \citep{yan14,zhang12}.
The shape of particle energy distribution could bring information on the  underlying acceleration processes (e.g. \citep{yan13,cao13,peng14,chen14}).
In the VHE band, Cherenkov telescopes cannot  regularly monitor AGNs, because of their limited duty
cycle and narrow field of view (FOV). Wide-FOV Extensive Air Showers (EAS) arrays, with high duty cycles, are more suitable for this purpose.
A review on EAS arrays and their observations of AGNs can be found in \citep{chen13}.

ARGO-YBJ is an EAS array with an energy threshold for primary $\gamma$-rays of $\sim$300 GeV. During 5 years ARGO-YBJ continuously monitored the blazar Mrk 421, extending at higher energies the multi-wavelength survey carried out by the Owens Valley Radio Observatory (OVRO), the satellite-borne X-ray detectors $Swift$, the Rossi X-ray Timing Explorer (RXTE), the Monitor of All-sky X-ray Image (MAXI) and the GeV $\gamma$-ray detector $Fermi$-LAT. In particular, thanks to the ARGO-YBJ and $Fermi$-LAT data, the high energy component of Mrk 421 SED has been completely covered without any gap from 100 MeV to 10 TeV. In this paper we report on the 4.5-year multi-wavelength data recorded from August 2008 to February 2013, a period that includes several large flares of Mrk 421. Such a long-term multi-wavelength observation is rare and provides a unique opportunity to investigate on the emission variability of Mrk 421 from radio frequencies to TeV $\gamma$-rays.

This work is organized as follows:In Section 2 we summarize the data at different wavelengths collected by
both satellite-borne and ground-based detectors. In Section 3 the light
curves and SEDs observed by the different detectors are presented. In
Section 4 the key parameters of the one-zone SSC emission model are
obtained by fitting the data, then the astrophysical implications   are discussed  in Section 5.  A summary is given in Section 6.  The cosmology parameters used in this paper are:  $H_0$ = 70 km $s^{-1}$ $Mpc^{-1}$, $\Omega_{M}$ = 0.3, and $\Omega_{\Lambda}$ = 0.7. The redshift of Mrk 421 corresponds to a luminosity distance
of 135.9 Mpc.

\section{Multi-wavelength observations and analysis}
The present section reviews the available data sets. We briefly summarize the energy range in which
each detector works and the data processing steps. More details can be found in the cited references.

\subsection{ARGO-YBJ VHE $\gamma$-ray data}
ARGO-YBJ is an EAS detector located at an altitude of 4300 m  a.s.l. (atmospheric depth 606 g cm$^{-2}$) at the Yangbajing Cosmic Ray Laboratory (30.11 N, 90.53 E) in Tibet, P.R. China. It is mainly devoted to   $\gamma$-ray astronomy
\citep{barto12a,barto12b,barto12c,barto13b,barto14,barto15}
and cosmic ray physics
\citep{barto12d,barto12e,barto13c}.
The detector consists of a carpet ($\sim$74$\times$ 78 m$^2$) of resistive plate chambers (RPCs) with $\sim$93\% of active area, surrounded by a partially instrumented area ($\sim$20\%) up to $\sim$100$\times$110 m$^2$.
Each RPC  is read out by 10 pads (55.6 cm$\times$61.8 cm) representing the space-time pixels of the detector.
Details of the detector layout  can be found in \citep{aielli06}.
Due to the full coverage configuration and the location at high altitude, the detector energy threshold is $\sim$300 GeV, much lower than any previous EAS array, as Tibet AS$\gamma$ \citep{ameno05} and Milagro \citep{{abdo14}}.

The ARGO-YBJ experiment,  with a $\sim$ 2 sr FOV, is able to monitor the sources in the sky
with a zenith angle less than 50$^{\circ}$.
At the ARGO-YBJ site,  Mrk 421 culminates with a zenith angle of 8.$^{\circ}$1,
and is observable for 8.1 (4.7) hours per day with a zenith angle
less than 50$^{\circ}$ (30$^{\circ}$).

The detector, in its full configuration, has been in stable data taking since November 2007  to February 2013,  with a 4\% of dead time  and an average duty cycle higher than $86\%$.
The detector performance and the analysis techniques
are extensively discussed in \cite{barto13a}.
The detector angular resolution depends on the number of triggered pads N$_{pad}$,
ranging from $1^{\circ}.7$  for N$_{pad} >$ 20 to $0^{\circ}.2$ for N$_{pad} >$ 1000.
The median primary energy of $\gamma$-rays is 0.36 TeV for events with N$_{pad} >$ 20
and 8.9 TeV for N$_{pad} >$ 1000 \citep{barto13a}.
The light curves presented here are obtained selecting the events with N$_{pad}>$60,
corresponding to a photon median energy of $\sim$1.1 TeV.
The cosmic ray background around the source direction is estimated using the
\emph{direct integral method} \citep{fley04}.
The   spectrum is estimated as described in \citep{barto11}, by comparing the detected  and the expected signal (i.e. the number of events)  as a function of N$_{pad}$.
Five intervals, N$_{pad}$=20-59, 60-99, 100-199, 200-499, and $>$500, are considered,
corresponding to a $\gamma$-ray energy range between 0.3 TeV and 10 TeV.
The spectrum is assumed to follow a single Power Law (PL): $f(E)=J_{0} \cdot E^{-\alpha}$.
The simulated events are sampled in the energy range from 10 GeV to 100 TeV.
The ARGO-YBJ detector response has been evaluated using a custom Montecarlo simulation (see \cite{guo10}).

\subsection{$Fermi$-LAT HE $\gamma$-ray data}
$Fermi$-LAT \citep{atwood09} is a pair-conversion telescope, with a FOV of over 2 sr,
operating in the energy range above 100 MeV.
$Fermi$-LAT started science data taking in August 2008. The data used in this work have been
downloaded through the Fermi science support center\footnote{http://fermi.gsfc.nasa.gov/ssc/}.
A circular region of $15^{\circ}$ radius centered on Mrk 421 was chosen for the event reconstruction.
The analysis was performed using the ScienceTool and the corresponding threads,
provided by the $Fermi$-LAT collaboration (Version v9r33p0)\footnote{http://fermi.gsfc.nasa.gov/ssc/}.
Events with zenith angles $<100^{\circ}$ were selected from the \emph{Source} class events,
which have the highest probability of being  photons.  The light curve was created through aperture
photometry, which allows a model independent determination of the flux, including both background and source emission.
To build the SED of the source we used the suggested \emph{gtlike} tool,
based on a binned maximum likelihood method. The instrument response function
is $P7REP_{-}SOURCE_{-}V15$, and the Galactic emission was reproduced using
the model of $gll_{-}iem_{-}v05_{-}rev1.fit$. The model for the
extragalactic isotropic diffuse emission was $iso_{-}source05_{-}rev1.txt$.
 All sources within $20^{\circ}$ from the Mrk 421 position
were taken into account. The spectral parameters are kept free for the sources within $10^{\circ}$,   while  are fixed to the values given in
the second $Fermi$-LAT catalogue \citep{nolan12} for other sources.

 To describe the source spectrum from 100 MeV to 500 GeV, we use two different models: the Power Law model and the LogParabolic model (LP). The latter is described by the expression  $f(E)=J_{0} \cdot (E/E_{0})^{-(a+b \cdot log(E/E_{0})}$, where   $E_{0}$ indicates the normalization energy,  $J_{0}$ the  flux at  $E_{0}$, $b$  the curvature around the SED peak and $a$  the spectral index below the SED peak. Following \cite{nolan12}, these two models are compared by defining the curvature test statistic TS$_{curve}$=(TS$_{LP}$ - TS$_{PL}$). The significance of the curvature can be approximately estimated as $\sqrt{TS_{curve}}$.

\subsection{$Swift$-BAT hard X-ray data}
$Swift$ Burst Alert Telescope (BAT)  \citep{gehre04} is a coded aperture mask imaging telescope (1.4 sr FOV)
operating since February 2005. It orbits the Earth every 1.5 hours and monitors the whole sky at
hard X-rays once per day. The daily flux from Mrk 421 at energy 15$-$50 keV is provided by $Swift$-BAT\footnote{Transient monitor results provided by the $Swift$-BAT team:  http://heasarc.gsfc.nasa.gov/docs/swift/results/ \\ transients/weak/.} \citep{krimm13}  and is used here to build the light curve.
To obtain the SED, we downloaded all the available data of Mrk 421 through the HEASARC data
archive\footnote{http://heasarc.nasa.gov/docs/archive.html}. The data analysis includes the recipes presented in \citep{ajello08,tueller10}. The spectrum was obtained as a weighted average of the source
spectra using a six-channel binning in the 14$-$195 keV energy range,
i.e., 14$-$22, 22$-$30, 30$-$47, 47$-$71, 71$-$121, and 121$-$195 KeV.
A power law function was used to fit the measured spectrum.

\subsection{$MAXI$-GSC  X-ray data}
The $MAXI$  Gas Slit Camera (GSC)   \citep{matsu09} detector, made of twelve one-dimensional position sensitive
proportional counters, operating in the 2$-$20 keV range, started data taking in August 2009.
The experiment achieves 97\% of sky coverage per day. The light curves for specific sources are
publicly available \footnote{The MAXI data are provided by RIKEN, JAXA and the MAXI team: http://maxi.riken.jp/top/} in three energy bands: 2$-$4, 4$-$10, and 10$-$20 keV.
These data were used in this work to build the X-ray spectrum of Mrk 421, comparing the measured
counting rate in each band with the one of the Crab Nebula, used as standard candle, as proposed by
the $Swift$-BAT collaboration \citep{tueller10}.

\subsection{$RXTE$-ASM soft X-ray data}
$RXTE$  All Sky Monitor (ASM) \citep{levine96} consists of three proportional counters, each one with a field of
view of $6^{\circ}\times90^{\circ}$.
It covers about 80\% of the sky during one full revolution in about 1.5 hr.
The $RXTE$-ASM data in the (2$-$12) keV range are publicly available\footnote{Quick-look results provided by the  $RXTE$-ASM team: http://xte.mit.edu/ASM\_lc.html.}. The light curves are given in
three energy bands: 1.5$-$3, 3$-$5, and 5$-$12 keV, which were used here to build the X-ray spectrum.
For Mrk 421 the daily flux is provided from 1995 up to the middle of 2010.

\subsection{$Swift$-XRT soft X-ray data}
$Swift$  X-ray Telescope (XRT)  \citep{burro05} is a focusing X-ray telescope with an energy range from 0.2 to 10 keV.
The light curves at 0.3$-$10 keV for Mrk 421 (available here\footnote{ http://www.swift.psu.edu/monitoring/}) were directly used in this work.
To obtain the SED all the $Swift$-XRT Windowed Timing (WT) Observations, available at HEASARC\footnote{http://heasarc.nasa.gov/docs/archive.html}, were downloaded. The $Swift$-XRT data set are calibrated using
the  calibration files   available in the Swift data base (CALDB\footnote{http://heasarc.gsfc.nasa.gov/docs/heasarc/caldb/caldb$_{-}$supported$_{-}$miss -ions.html}) and processed with the XRTDAS software package (distributed
by HEASARC within the HEASoft package (v.6.16)\footnote{http://heasarc.nasa.gov/lheasoft/}) using
the \emph{xrtpipeline} task.
 Events for the spectral analysis were selected within a circle
of 30 pixels ($~$71$^{\prime\prime}$) radius  centered on the source position.
The background was extracted from an annular
region with a 40 pixels inner radius and 80 pixels outer radius, also centered on the source position.  The count rate is less than 100 Hz for all the observations considered in this work, so the WT mode data should not be affected by pile-up effects.
The average spectrum in the 0.3$-$10 keV energy
band was fitted using the XSPEC package \footnote{http://heasarc.gsfc.nasa.gov/xanadu/xspec/} (v.12.8.2), assuming a LP model (fixing $E_{0}$=1 keV),   with  an absorption hydrogen-equivalent column
density set to the Galactic value in the direction of the source, namely $1.92 \times 10^{20}$ $cm^{-2}$ \citep{kalber05}.
 For such a spectrum model, the energy of the SED peak is estimated as $E_{peak}=10^{(2-a)/2b}$.
In addition, a small energy offset ($\sim$40 eV) was applied to the observed energy spectrum, according to
\cite{abdo11}.

\subsection{$Swift$-UVOT ultraviolet data}
$Swift$  Ultraviolet/Optical Telescope (UVOT)  \citep{roming05} is the ultraviolet and optical Telescope onboard the satellite. All
the $Swift$-UVOT observations of Mrk 421, at the three ultraviolet bands (UVW1, UVM2 and UVW2)
available at the HEASARC data archive, were included in our analysis.  The level 2  UVOT images from the archive,
produced by a custom UVOT pipeline with data screening and coordinate transformation, were directly used in this work.
The photometry was
computed using a 8$^{\prime\prime}$ source region centered on the Mrk 421 position,
performing the calibrations presented in \cite{poole08}, which  also convert UVOT magnitudes to flux units. The latest UVOT calibration files released on January 18th 2013 were used here.
The background was extracted from an annular region (with radius of 20$^{\prime\prime}$ $-$ 50$^{\prime\prime}$)  centered on the source position. The
flux has been corrected for the Galactic extinction using the \cite{card89} parameterization,
with  E$_{B-V}$ = 0.013 mag \citep{schlaf11}.

\subsection{OVRO radio data}
The OVRO \citep{richard11}
is a 40-m radio telescope working at 15.0 GHz with 3 GHz bandwidth. Mrk 421 was observed by OVRO as part of the blazar monitoring programme, which
observed a sample of over 1800 AGNs twice per week.
Mrk 421 has been included since the end of 2007.
The light curve for Mrk 421 is
publicly available \footnote{http://www.astro.caltech.edu/ovroblazars/} and is directly used in this work.
The systematic error is estimated to be about 5 percent of the flux density, which is not included in the error bars.

\section{Results}
Figure 1 summarizes the temporal and energy coverage of
the different instruments considered in this analysis.
To monitor efficiently the HE and VHE components of Mrk 421 spectrum,
we limit the observation time to the $\sim$4.5 years in which
the data of $Fermi$-LAT (100 MeV$<$E$<$500 GeV) and ARGO-YBJ (E$>$300 GeV) overlap,
i.e. since 2008 August 5 (start time of $Fermi$-LAT science data acquisition)
to 2013 February 7 (end time of ARGO-YBJ data taking).

\begin{figure}
\includegraphics[width=3.5in,height=2.4in]{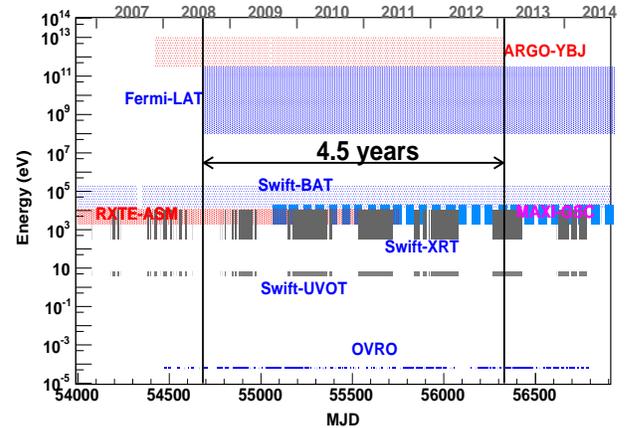}
\caption{
Time and energy coverage of different detectors in 4.5 years
of Mrk 421 observation.
}
\label{fig1}
\vspace*{0.5cm}
\end{figure}

In the following, we firstly determine the light curves of Mrk 421
in the energy ranges explored by the different detectors.
Then, by inspecting the light curves, we define the flaring and steady phases of the source.
Finally, we analyze the general features of the corresponding SEDs.

\subsection{Light curves}
Figure 2 shows the light curves of Mrk 421, as obtained by the data of the previously described
 experiments, covering the entire energy range from radio  to the TeV band.
The time integration is chosen taking into account the sensitivity of the instruments.
For ARGO-YBJ each point corresponds to one month (30 days) of data,
while for $Fermi$-LAT, $Swift$-BAT, $RXTE$-ASM and $MAXI$-GSC the data are averaged over one
week. For $Swift$-XRT and $Swift$-UVOT each point is the result of each dwell, which last about hundreds
of seconds. Note that since the $Swift$-UVOT light curves in the three photometric bands (UVW1, UVM2, and
UVW2) show similar behaviors, only the light curve of UVW1 is considered here.
 For OVRO each point is the result of each observation.  The data presented in Figure 2 are also listed in Table 1.

\begin{figure*}
\includegraphics[width=7in,height=1.4in]{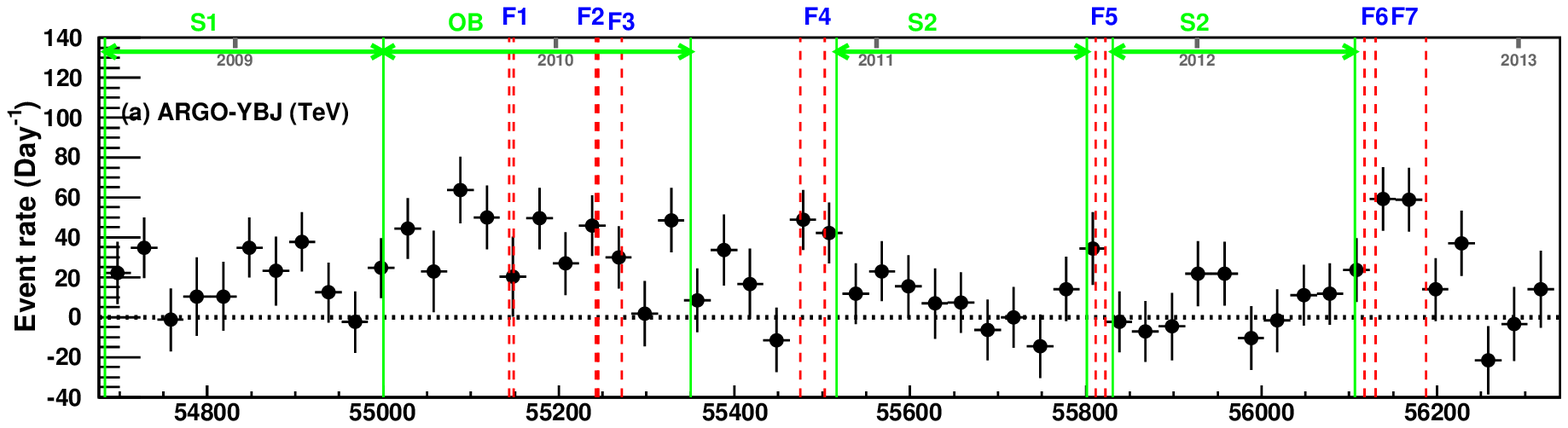}
\includegraphics[width=7in,height=1.4in]{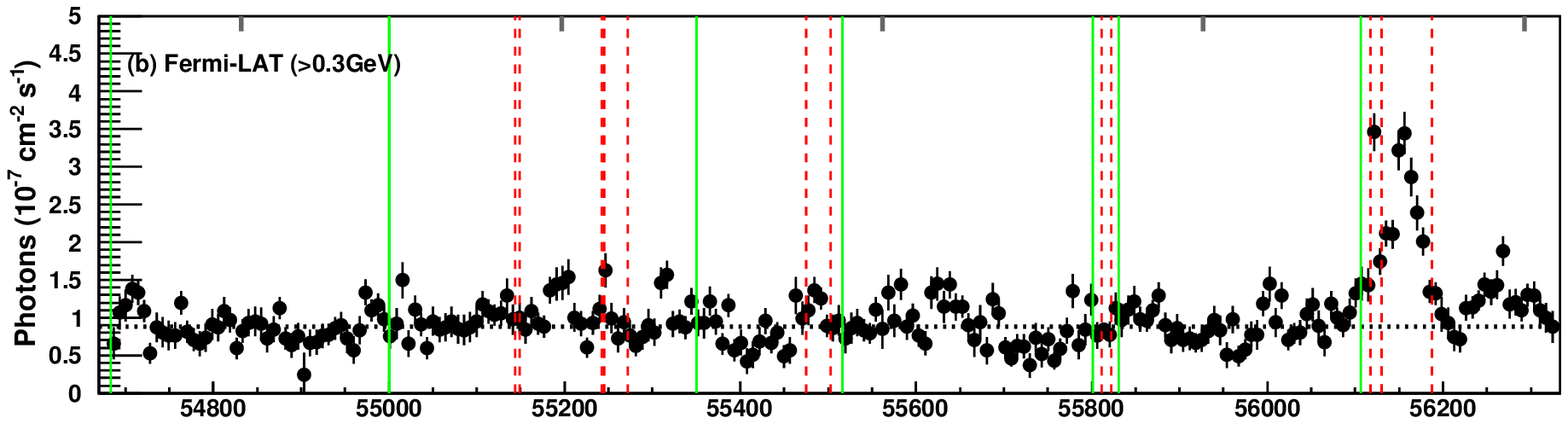}
\includegraphics[width=7in,height=1.4in]{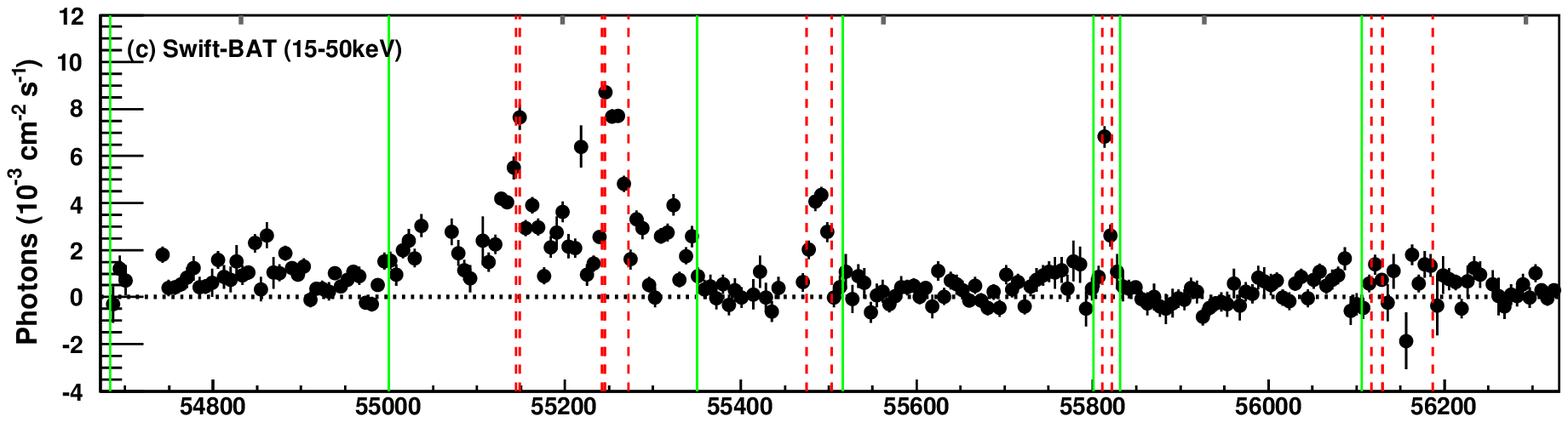}
\includegraphics[width=7in,height=1.4in]{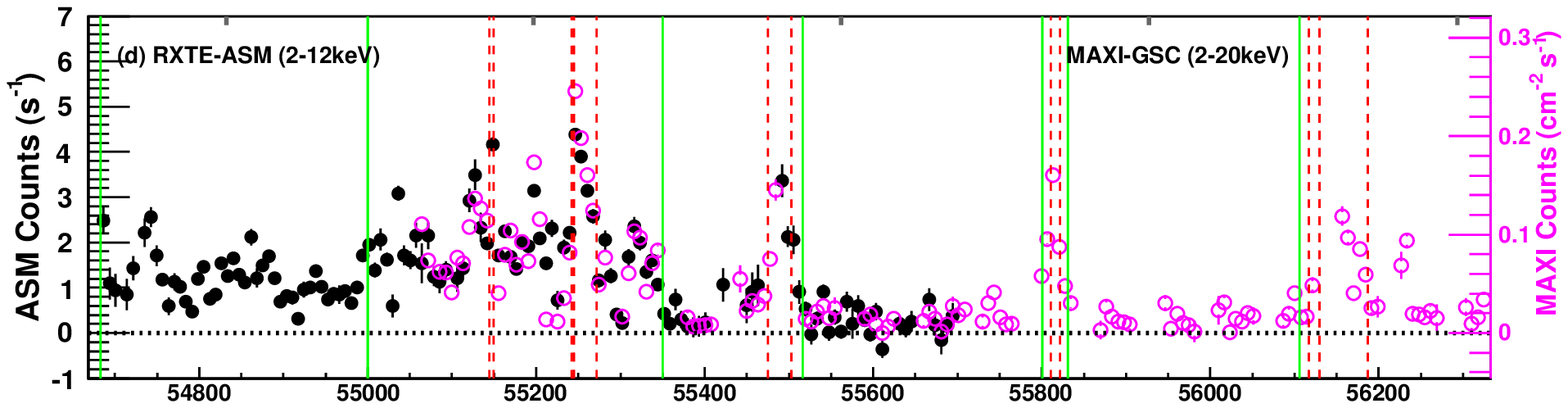}
\includegraphics[width=7in,height=1.4in]{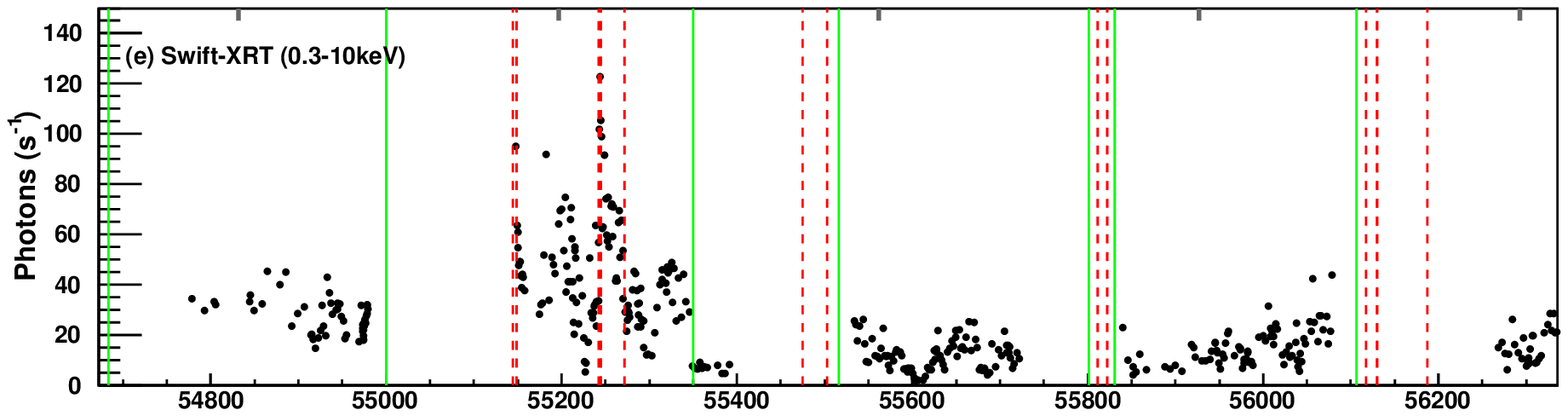}
\includegraphics[width=7in,height=1.4in]{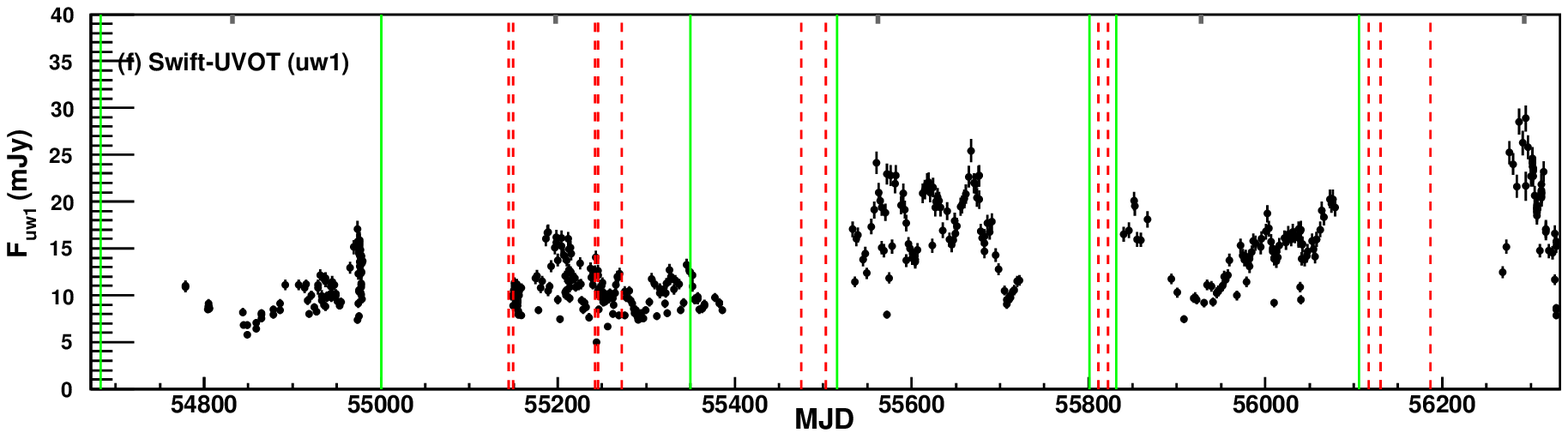}
\includegraphics[width=7in,height=1.4in]{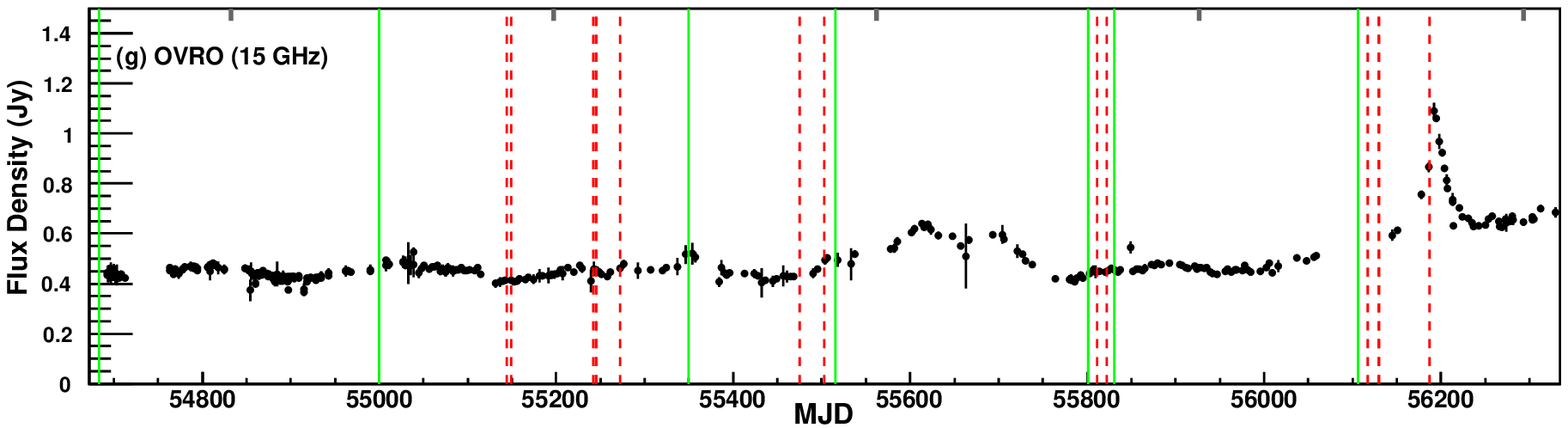}
\caption{
Mrk 421 light curves in different energy bands, from 2008 August 5 to 2013 February 7.
Each bin of the ARGO-YBJ data contains the event rate averaged over 30 days.
Each bin of the  $Fermi$-LAT, $Swift$-BAT, $RXTE$-ASM and $MAXI$-GSC data
contains the event rate averaged over 7 days.  The horizontal dotted line in panel (b) indicates the average flux before F6 and the horizontal dotted lines in the other panels indicate the zero flux.
}
\label{fig2}
\vspace*{0.5cm}
\end{figure*}

\begin{table}
\centering
\caption{Light curves shown in Figure 2}
\label{tab-1}
\vspace{2mm}
\begin{tabular}{ccccc}
\hline
 MJD & $\bigtriangleup$T (Day)  & Flux$^a$  & $\bigtriangleup$Flux$^a$ & Detector \\
\hline
54698.00 & 15.00 & 2.231e+01 & 1.560e+01 & ARGO-YBJ \\
54728.00 & 15.00 & 3.486e+01 & 1.527e+01 & ARGO-YBJ \\
54758.00 & 15.00 & -1.327e+00 & 1.570e+01 & ARGO-YBJ \\
54788.00 & 15.00 & 1.026e+01 & 1.973e+01 & ARGO-YBJ \\
54818.00 & 15.00 & 1.047e+01 & 1.737e+01 & ARGO-YBJ \\
54848.00 & 15.00 & 3.495e+01 & 1.492e+01 & ARGO-YBJ \\
54878.00 & 15.00 & 2.322e+01 & 1.723e+01 & ARGO-YBJ \\
54908.00 & 15.00 & 3.786e+01 & 1.477e+01 & ARGO-YBJ \\
54938.00 & 15.00 & 1.242e+01 & 1.493e+01 & ARGO-YBJ \\
54968.00 & 15.00 & -2.462e+00 & 1.544e+01 & ARGO-YBJ \\
54998.00 & 15.00 & 2.471e+01 & 1.505e+01 & ARGO-YBJ \\
55028.00 & 15.00 & 4.428e+01 & 1.520e+01 & ARGO-YBJ \\
55058.00 & 15.00 & 2.294e+01 & 2.037e+01 & ARGO-YBJ \\
55088.00 & 15.00 & 6.377e+01 & 1.677e+01 & ARGO-YBJ \\
55118.00 & 15.00 & 4.981e+01 & 1.594e+01 & ARGO-YBJ \\
55148.00 & 15.00 & 2.032e+01 & 2.004e+01 & ARGO-YBJ \\
55178.00 & 15.00 & 4.956e+01 & 1.539e+01 & ARGO-YBJ \\
55208.00 & 15.00 & 2.683e+01 & 1.561e+01 & ARGO-YBJ \\
55238.00 & 15.00 & 4.596e+01 & 1.528e+01 & ARGO-YBJ \\
55268.00 & 15.00 & 2.997e+01 & 1.557e+01 & ARGO-YBJ \\
\hline
\multicolumn{5}{l}{$^a$ The flux units are events day$^{-1}$ for ARGO-YBJ, photons cm$^{-2}$ s$^{-1}$ for  }\\
\multicolumn{5}{l}{\quad  $Fermi$-LAT, $Swift$-BAT and $MAXI$-GSC, photons  s$^{-1}$ for $RXTE$-ASM}\\
\multicolumn{5}{l}{\quad  and $Swift$-XRT, mJy for $Swift$-UVOT, and Jy for OVRO. }\\
\multicolumn{5}{l}{This table is available in its entirety in a machine-readable form in the }\\
\multicolumn{5}{l}{online journal.   A portion is shown here for guidance regarding its}\\
\multicolumn{5}{l}{form and content.}
\end{tabular}
\vspace*{0.5cm}
\end{table}

According to the long-term  light curves presented in Figure 2, Mrk 421 showed both low and high activity phases at all wavebands during the 4.5 years considered in this work.
To quantify the variability amplitudes in each energy band, the normalized variability amplitude (F$_{var}$), defined according to \citep{edelson96}, was computed as
\begin{equation}
F_{var}=\frac{\sqrt{\sigma_{tot}^2-\sigma_{err}^2}}{\bar{F}}
\end{equation}
where $\sigma_{tot}$ is the standard deviation of the flux, $\sigma_{err}$ is the mean error of the flux points, and $\bar{F}$ is the mean flux.
To facilitate the comparison of   F$_{var}$ for different bands, we rebinned  all light curves shown in Figure 2
with the same bin size, i.e., 7 days per bin.
 The ARGO-YBJ data with 7-day bin size are presented in Table 1.
 The  F$_{var}$ as a function of band energy  is shown in Figure 3. The variability  amplitude
increases from 21\% in radio   to 137\% in hard X-rays. The amplitude is 39\% for   GeV $\gamma$-rays, and it increases
to 84\% at TeV energies.
It should be noted that the light curve of GeV $\gamma$-rays is obtained through the aperture
photometry method, which includes a contribution from  the background  emission at a 18\% level according to our estimation.
Since the background   only affects the average flux and not the variability
amplitude, the effective amplitude of   GeV $\gamma$-ray variability is 47\%.

\begin{figure}
\includegraphics[width=3.5in,height=2.4in]{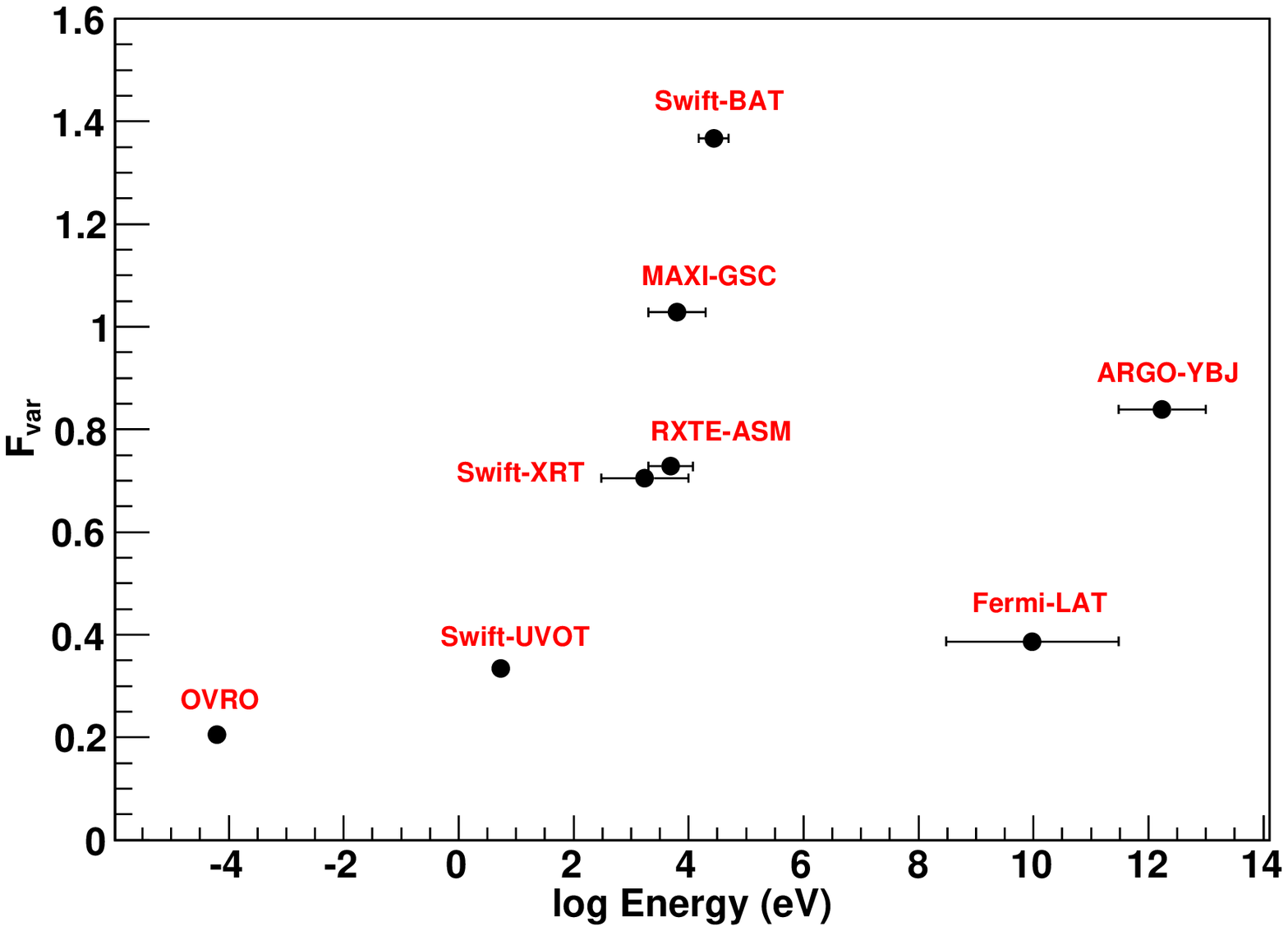}
\caption{
The normalized variability amplitude F$_{var}$ for different energy bands.
}
\label{fig3}
\vspace*{0.5cm}
\end{figure}

%\begin{table*}
\begin{table}
\centering
\caption{Correlation coefficient}
\label{tab-1}
\vspace{2mm}
\begin{tabular}{c|ccc}
\hline
  & r$_{UVOT}^a$ & r$_{BAT}^b$ & r$_{LAT}^c$ \\
\hline
ORVO         & 0.62$\pm$0.14$^d$  & -0.26$\pm$0.06  &0.38$\pm$0.11$^e$    \\
$Swift$-UVOT &  ...            &-0.35$\pm$0.08  &0.33$\pm$0.11   \\
$Swift$-XRT  &  -0.39$\pm$0.08 & 0.85$\pm$0.21  &0.27$\pm$0.15   \\
$RXTE$-ASM   &  -0.47$\pm$0.10 & 0.87$\pm$0.17  &0.38$\pm$0.14   \\
$MAXI$-GSC   &  -0.34$\pm$0.12 & 0.89$\pm$0.22  &0.30$\pm$0.10  \\
$Swift$-BAT  &  -0.35$\pm$0.08 & ...              &-0.01$\pm$0.08  \\
$Fermi$-LAT  &  0.33$\pm$0.11  &-0.01$\pm$0.08  &...   \\
ARGO-YBJ     &  -0.90$\pm$0.33 & 0.76$\pm$0.25  &0.61$\pm$0.22   \\
\hline
\multicolumn{4}{l}{$^a$ Cross-correlation coefficient with  $Swift$-UVOT.}  \\
\multicolumn{4}{l}{$^b$ Cross-correlation coefficient with  $Swift$-BAT.}  \\
\multicolumn{4}{l}{$^c$  Cross-correlation coefficient with  $Fermi$-LAT. }\\
\multicolumn{4}{l}{$^d$  The coefficient at lag=21 day is $r$=0.79$\pm$0.17. }\\
\multicolumn{4}{l}{$^e$  The coefficient at lag=42 day is $r$=0.83$\pm$0.27. }\\
 \end{tabular}
\vspace*{0.5cm}
 \end{table}

According to  Figure 2, only the light curves of $Swift$-BAT, $Fermi$-LAT and ARGO-YBJ  continuously  sampled  the whole 4.5 years period considered here. Several large X-ray and GeV $\gamma$-ray flares are  visible from the light curves of $Swift$-BAT and $Fermi$-LAT. The flux variability in the $Swift$-BAT and $Fermi$-LAT energy bands seems not to be correlated during   flares.  The variability of VHE $\gamma$-ray flux is roughly correlated both with  X-ray and GeV flares. The variability of radio and UV flux seems not to be correlated with that of X-rays and $\gamma$-rays. To be more rigorous, the discrete correlation function (DCF; \citep{edelson88}) is used to quantify the degree of correlation  between the light curve of $Swift$-BAT ($Fermi$-LAT, $Swift$-UVOT) with the others. To uniform the data,  the cross-correlation analysis was performed, using  weekly binned light curves. No significant time lag (within [-200,200] days)  was measured in this analysis except between $Fermi$-LAT and ORVO, where it was found that the GeV $\gamma$-rays lead  the radio by   42 days, with a correlation coefficient  $r$=0.83$\pm$0.27.  This result is consistent with \citep{hovatta15} who also measured a 40$\pm$9 days time lag using 4 years ORVO and $Fermi$-LAT data. Beside this,
a possible time lag is measured between $Swift$-UVOT and ORVO data.
The UV flux seems to lead the radio by 21 days with a correlation coefficient $r$=0.79$\pm$0.17, that however   is comparable to the coefficient $r$=0.62$\pm$0.14 obtained for a time lag equal to zero.
The correlation coefficients for a time lag  zero are listed in Table 2, for all the datasets.  According to our analysis,
the $Swift$-BAT hard X-ray flux is weakly anti-correlated with the radio and UV flux,  while is significantly correlated with the soft X-ray flux, not correlated with GeV $\gamma$-rays and clearly correlated with VHE $\gamma$-rays. The $Fermi$-LAT GeV $\gamma$-ray flux is weakly correlated with radio, UV and soft X-rays, and moderately correlated with VHE $\gamma$-rays. The UV flux appears to be moderately correlated with radio, weakly anti-correlated with X-rays, and clearly anti-correlated with VHE $\gamma$-rays.  It should be noted however   that  the observation time of $Swift$-UVOT has several long gaps, which  would affect the cross-correlation analysis. In particular, the  anti-correlation with   VHE $\gamma$-rays needs to be checked by future observations.

\subsection{Source state definition}
In this paper  we will focus on the large X-ray and GeV $\gamma$-ray flares, with the aim  to investigate the spectral variation at different wavebands, compared to the low activity states. We will define different states of activity for Mrk 421 mainly basing on the light curves of $Fermi$-LAT and $Swift$-BAT, partially taking into account the curves of $RXTE$-ASM and $MAXI$-GSC.
For X-ray flares, we will only select the flares which show a large increase both in hard and soft X-rays.

From August 2008  (MJD=54683)  to June 2009  (MJD=55000),  Mrk 421 shows a low activity at all wavebands.
We mark this period as Steady 1 (S1) phase. It
should be noted that, during this period, a 4.5 months long multi-frequency campaign was
organized \citep{abdo11}.  Afterwards, according to the X-ray light curves of
$Swift$-BAT, $RXTE$-ASM and $MAXI$-GSC, the source entered a long-lasting outburst phase
starting in June 2009  and ending in June 2010  (MJD=55350), which we denoted as Outburst (OB).
The X-ray flux is higher than in the S1 period and also varies with time.
During this active phase, three large flares, named Flare 1, Flare 2 and
Flare 3 (F1, F2 and F3) are clearly detected both by $Swift$-BAT and $RXTE$-ASM.
During F1, the flux starts to increase on 2009 November 9 (MJD=55144), reaches the maximum on November 12,
then decreases to a quasi-steady state on November 14. The average flux is about 3 and 14
times higher than in the S1 phase, in the 2$-$12 keV and 15$-$50 keV ranges, respectively.
The F2 (from 2010 February 15 (MJD=55242) to 17) and F3 (from 2010 February 18  to March 16 (MJD=55271)) states were
reported by MAXI at 2$-$10 keV \citep{isobe10}. A zoom view of the light curves during both flares is shown in Figure 4, with a 3-day binning for ARGO-YBJ and one-day binning for the other experiments.
F2 is a very fast flare reaching the peak flux in one day and then decaying in one day too. This flare
is associated with the huge VHE $\gamma$-ray  flare with a flux around 10 times the Crab Nebula flux
(I$_{crab}$)  detected by VERITAS on February 17 (MJD=55244) \citep{ong10}. The $\gamma$-ray flux
enhancement is also evident in $Fermi$-LAT and ARGO-YBJ data. F3 follows flare 2.
Note that for the OB state, the embedded durations of
flares F1, F2 and F3 are excluded.

\begin{figure*}
\includegraphics[width=7in,height=7.in]{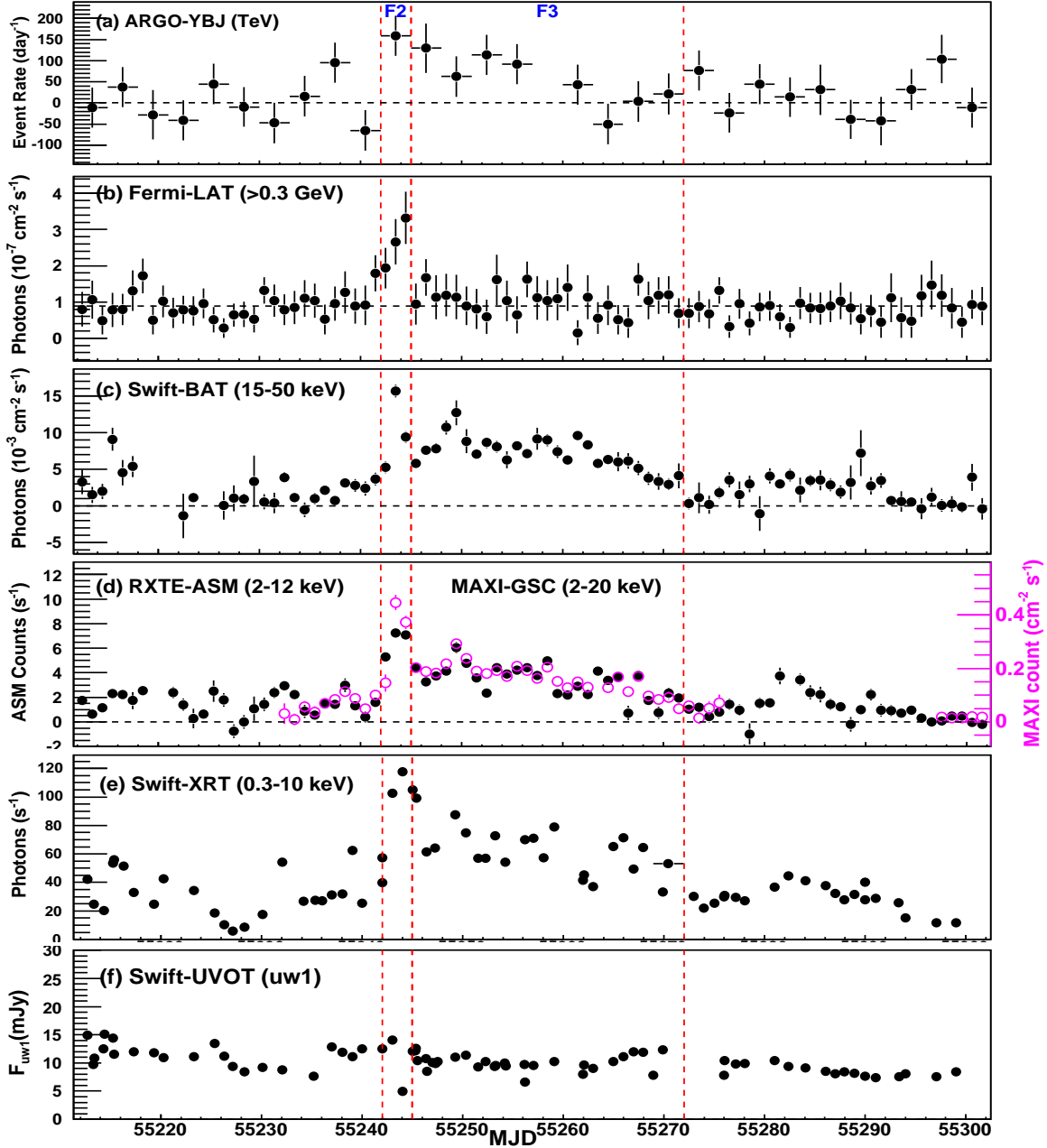}
\caption{
Mrk 421 light curve in different energy bands, from 30 days before Flare 2 to 30 days after Flare 3.
}
\label{fig4}
\vspace*{0.5cm}
\end{figure*}

After May 2010, Mrk 421 entered a low steady phase that ended on October 6 (MJD=55475), when the
flux of both hard and soft X-rays started gradually to increase for two weeks. The whole period
lasted about one month and is marked as Flare 4 (F4). Then Mrk 421 came to a long and
steady phase (S2), both in X-rays and $\gamma$-rays, which lasted about 1.6 years,
since November 2010 (MJD=55516) to June 2012 (MJD=56106).
This is the longest steady phase during the monitored period, therefore it has been
picked out as a baseline reference to all the other states. The embedded strong flare
denoted as Flare 5 (F5), occurred in September 2011 (MJD=55811) and lasting $\sim$7 days,
has been excluded from S2.

In the whole year 2012 the flux in hard X-rays was almost stable, while the GeV $\gamma$-ray flux
measured by $Fermi$-LAT entered into a high flux level from 2012 July 9 (MJD=56117) to September 17 (MJD=56187).
This is the first long-term GeV $\gamma$-ray flare from Mrk 421 ever detected,
reported by both $Fermi$-LAT \citep{amman12}  and ARGO-YBJ  \citep{barto12f}.
According to the GeV $\gamma$-ray light curve, two peaks are selected, marked as Flare 6
(F6, from 2012 July 9 to 21 (MJD=56129) and Flare 7 (F7, from 2012 July 22  to September 16),
during which also the VHE $\gamma$-ray flux detected by ARGO-YBJ seems to be enhanced.
Some hints of enhancement were partly observed by $MAXI$-GSC in the soft X-ray energy range.

The light curves during F1, F4, F5, F6, F7 are shown in Figure 5, where the flux measured by ARGO-YBJ is averaged over the different flare durations, and a 3-day binning is used for the other experiments.
The duration of all the selected states are summarized in Table 3.
Note that $Swift$-XRT data are only available for F1, F2, F3, S1, S2 and OB states.

\begin{figure*}
\includegraphics[width=7.02in,height=1.4in]{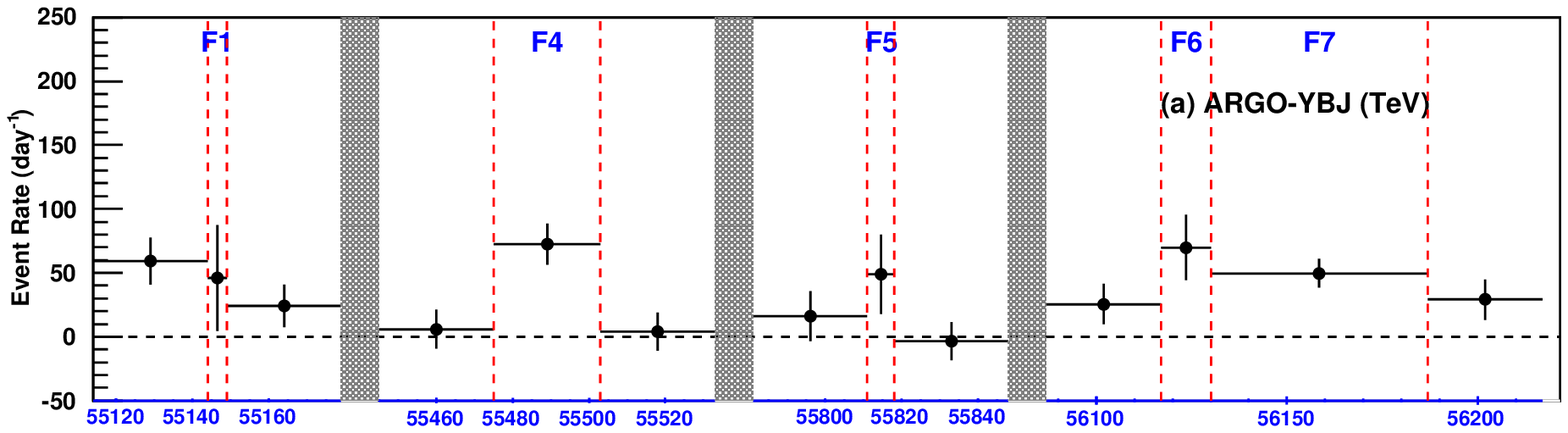}
\includegraphics[width=7in,height=1.4in]{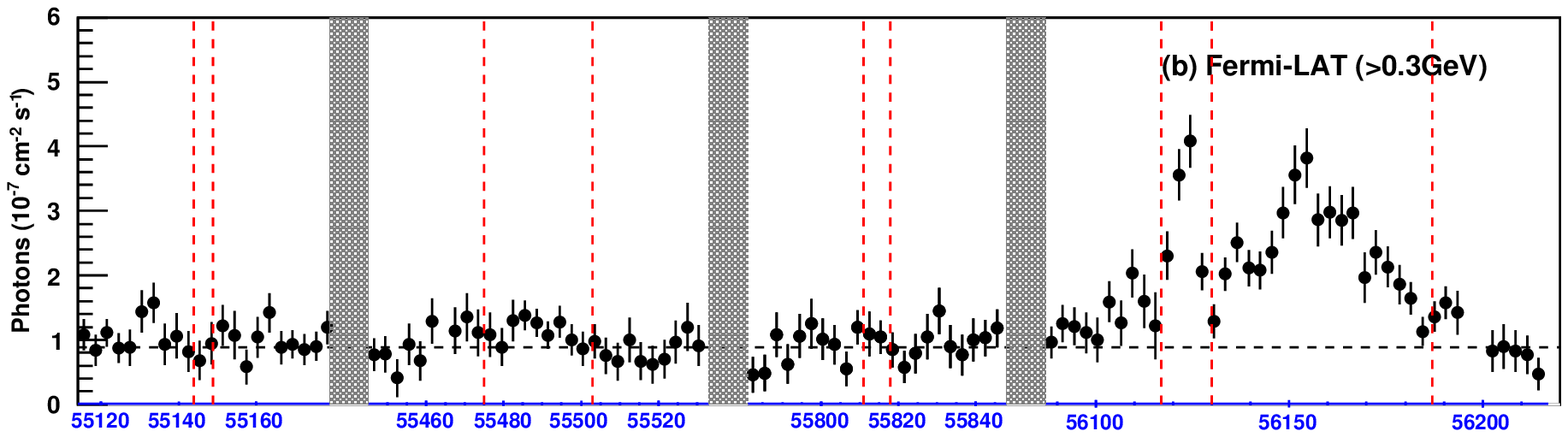}
\includegraphics[width=7in,height=1.4in]{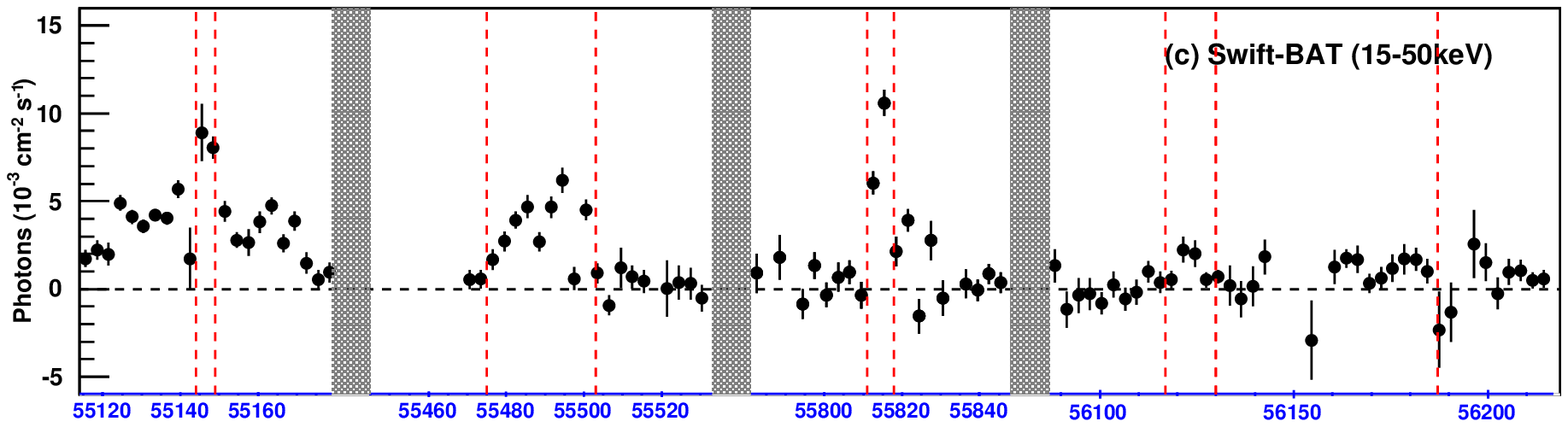}
\includegraphics[width=7in,height=1.4in]{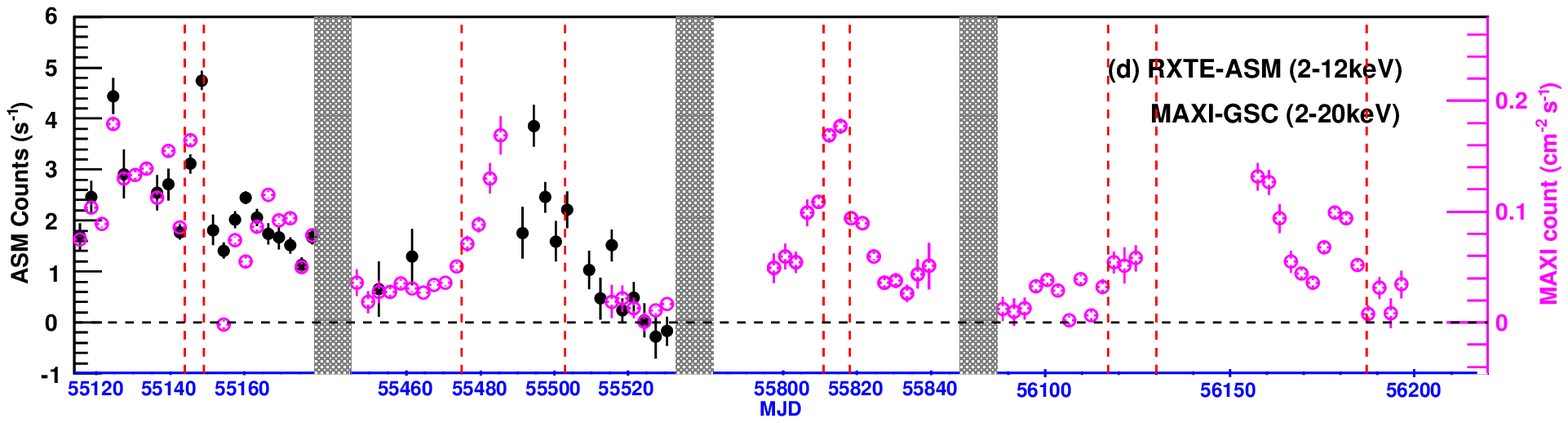}
%\plotone{Fig5a.eps} \plotone{Fig5b.eps} \plotone{Fig5c.eps} \plotone{Fig5d.eps}
\caption{
Mrk 421 light curve in different energy bands, from 30 days before to 30 days after the five flares.
Each bin of $Fermi$-LAT, $Swift$-BAT, $RXTE$-ASM, $MAXI$-GSC contains the event rate averaged over 3 days.
}
\label{fig5}
\vspace*{0.5cm}
\end{figure*}

\begin{table*}
\centering
\caption{SEDs of Mrk 421 during 10 states (the unit of the integral flux F is photons cm$^{-2}$ s$^{-1}$)}
\label{tab-1}
\vspace{2mm}
\begin{tabular}{c|c|cc|cc|cc|cc}
\hline
State & MJD& F$_{2-20keV}$   & $\alpha$ & F$_{14-195keV}$  & $\alpha$ & F$_{0.1-500GeV}$    & $\alpha$ & F$_{>1TeV}$      & $\alpha$\\
            &    & ($\times 10^{-2}$) &      & ($\times 10^{-3}$) &          & ($\times 10^{-7}$) &          & (I$_{crab}^{a}$) & \\
\hline
Flare 1 &55144$-$55149 & 18.3$\pm$1.8  &2.00$\pm$0.16 & 23.1$\pm$2.6     &2.78$\pm$0.27    &1.11$\pm$0.36 &1.57$\pm$0.15    &3.2$\pm$0.9 &2.78$\pm$0.36\\
Flare 2 &55242$-$55245 & 41.4$\pm$4.2  &2.38$\pm$0.15 & 19.2$\pm$1.9     &2.76$\pm$0.18    &3.48$\pm$0.60 &1.50$\pm$0.08    &7.2$\pm$1.5 &2.61$\pm$0.27\\
Flare 3 &55245$-$55272 & 19.9$\pm$1.6  &2.20$\pm$0.11 & 12.9$\pm$0.9     &2.61$\pm$0.13    &1.84$\pm$0.21 &1.79$\pm$0.09    &1.4$\pm$0.5 &2.42$\pm$0.46\\
Flare 4 &55475$-$55503 & 10.0$\pm$1.1  &2.17$\pm$0.22 & 5.82$\pm$0.83    &3.02$\pm$0.30    &2.16$\pm$0.19 &1.74$\pm$0.05    &1.9$\pm$0.5 &2.85$\pm$0.26\\
Flare 5 &55811$-$55818 & 17.0$\pm$1.6  &2.13$\pm$0.15 & 13.3$\pm$1.4     &2.74$\pm$0.23    &2.23$\pm$0.33 &1.79$\pm$0.09    &2.1$\pm$0.8 &2.75$^{b}$ \\
Flare 6 &56117$-$56130 & 6.15$\pm$0.96 &2.05$\pm$0.35 & 1.45$\pm$0.40    &2.47$\pm$0.62    &6.05$\pm$0.39 &1.68$\pm$0.04    &1.7$\pm$0.6 &2.84$\pm$0.39 \\
Flare 7 &56130$-$56187 & 8.70$\pm$0.91 &2.97$\pm$0.18 & 1.53$\pm$0.36    &3.09$\pm$0.61    &5.20$\pm$0.19 &1.75$\pm$0.02    &1.1$\pm$0.4 &3.22$\pm$0.24 \\

Outburst&55000$-$55350$^{c}$ &8.91$\pm$0.74 &2.41$\pm$0.11 & 4.58$\pm$0.59    &2.97$\pm$0.23   &1.92$\pm$0.06 &1.76$\pm$0.02    &0.91$\pm$0.14 &2.67$\pm$0.16 \\

Steady 1& 54683$-$55000& ...   & ... & 1.50$\pm$0.12   &2.51$\pm$0.16   &1.53$\pm$0.05 &1.75$\pm$0.02     &0.56$\pm$0.13 &2.64$\pm$0.27\\
Steady 2& 55516$-$56106$^{d}$&  1.78$\pm$0.18  &  2.38$\pm$0.19  &  0.318$\pm$0.087  & 3.0$^{b}$  & 1.69$\pm$0.04$^{e}$  & 1.77$\pm$0.01     & 0.33$\pm$0.10 & 2.75$^{b}$\\
\hline
Detector & & \multicolumn{2}{c|}{$MAXI$-GCS} & \multicolumn{2}{c|}{$Swift$-BAT} & \multicolumn{2}{c|}{$Fermi$-LAT} &\multicolumn{2}{c}{ARGO-YBJ} \\
\hline
\multicolumn{10}{l}{$^a$ I$_{crab}=1.85\times 10^{-11}$.}  \\
\multicolumn{10}{l}{$^b$  The spectral index is fixed.  }\\
\multicolumn{10}{l}{$^c$  The periods of Flare 1, 2 and 3 have been excluded. }\\
\multicolumn{10}{l}{$^d$  The periods of MJD 55801-55831 including Flare 5  have been excluded. }\\
\multicolumn{10}{l}{$^e$  The flux is 1.55$\pm$0.04 assuming a logparabolic spectrum model. }\\
 \end{tabular}
\vspace*{0.5cm}
 \end{table*}

\subsection{Spectral Energy Distribution}

In this section we report the multi-wavelength SEDs observed by the running experiments
in the outlined states.

To model the spectral energy distribution f(E), we assume a simple power law spectrum
for $Swift$-BAT, $RXTE$-ASM, $MAXI$-GSC, $Fermi$-LAT and ARGO-YBJ, while for $Swift$-XRT we assume
a logparabolic function. In the following, F represents the integral flux over the detector energy range, $\alpha$ the spectral index of the power law function, while $a$ and $b$ the parameters
of the logparabolic model (see Section 2 for details).
During S2, the $Swift$-BAT data are not significant enough to fit both the flux and the spectral index.
For such a situation, the spectral index is  fixed to 3.0.
A similar assumption has been chosen for ARGO-YBJ data by
fixing the spectral index to 2.75 during F5 and S2.
The time-averaged SEDs for the different activity states, obtained by fitting the data of all the
experiments, are summarized in Table 3 and 4. The flux  at each energy is   shown in Figure 7 and also listed in Table 5.
Note that in the following text, the first and the second components refer to the two SED bumps of the SED, as foreseen in the SSC model.

\subsubsection{$Swift$-XRT SED}

\begin{table*}
\centering
\caption{SEDs of Mrk 421 during 7 states (the unit for integral flux F is photons cm$^{-2}$ s$^{-1}$)}
\label{tab-1}
\vspace{2mm}
\begin{tabular}{c|cccc|cc}
\hline
 State& F$_{0.3-10keV}$                        & a    & b   & E$_{peak}$& F$_{2-12keV}$     & $\alpha$  \\
            &   &      &  &(keV)  &  ($\times 10^{-2}$)   & \\
\hline
Flare 1 &  1.227$\pm$0.003  &1.420$\pm$0.004 & 0.412$\pm$0.008   & 5.06$\pm$0.18       &18.40$\pm$0.73 &1.86$\pm$0.08\\
Flare 2 &  1.251$\pm$0.004  &1.656$\pm$0.004 & 0.390$\pm$0.009   & 2.76$\pm$0.07       &27.89$\pm$0.76 &2.02$\pm$0.06\\
Flare 3 &  0.903$\pm$0.001  &1.690$\pm$0.002 & 0.393$\pm$0.003   & 2.48$\pm$0.02       &13.74$\pm$0.29 &1.89$\pm$0.04\\
Flare 4 &   ...               &    ...           &    ...            &                      &10.15$\pm$0.67 &1.65$\pm$0.12\\
Outburst&  0.5746$\pm$0.0003  &1.864$\pm$0.001 & 0.448$\pm$0.002 & 1.419$\pm$0.004         &7.19$\pm$0.13 &2.19$\pm$0.04\\
Steady 1&  0.4526$\pm$0.0004  &2.104$\pm$0.001 & 0.462$\pm$0.003 & 0.771$\pm$0.003         &4.86$\pm$0.11 &2.41$\pm$0.06\\
Steady 2&  0.229$\pm$0.0003  &2.352$\pm$0.001 & 0.434$\pm$0.003 & 0.394$\pm$0.003         &0.68$\pm$0.18 &2.16$\pm$0.49 \\
\hline
Detector & \multicolumn{4}{c|}{$Swift$-XRT} & \multicolumn{2}{c}{$RXTE$-ASM}  \\
\hline
 \end{tabular}
\vspace*{0.5cm}
 \end{table*}
According to the $Swift$-XRT data at 0.3$-$10 keV, the peak energy of the
first component $E_{peak}$ is  0.394$\pm$0.003 keV and 0.771$\pm$0.003 keV during S2 and S1, respectively.
It increases to 1.429$\pm$0.004 keV during the OB phase and even up to 2.4$-$5.1 keV during F1, F2, and
F3.  The  correlation between the flux and $E_{peak}$ is shown in Table 4 and Figure 6. A power-law function is adopted to fit their relation, yielding  $f(E_{peak})=(0.2056\pm0.004) \cdot E_{peak}^{-1.266\pm 0.004}$  keV$^{-1}$ cm$^{-2}$ s$^{-1}$ with $\chi^{2}/dof$=2113/4.
The S1, S2, OB, F3 roughly follow  this function, while F1 and F2 clearly deviate,  indicating a different behavior of F1 and F2 with respect to the other flaring states.
It is worth to note that the $Swift$-XRT observations during Flare 1 only cover the period
with the maximum flux. For this reason, the measured flux is higher than those of $RXTE$-ASM and $MAXI$-GSC, as shown in Figure 7.

\begin{figure}
\includegraphics[width=3.5in,height=2.4in]{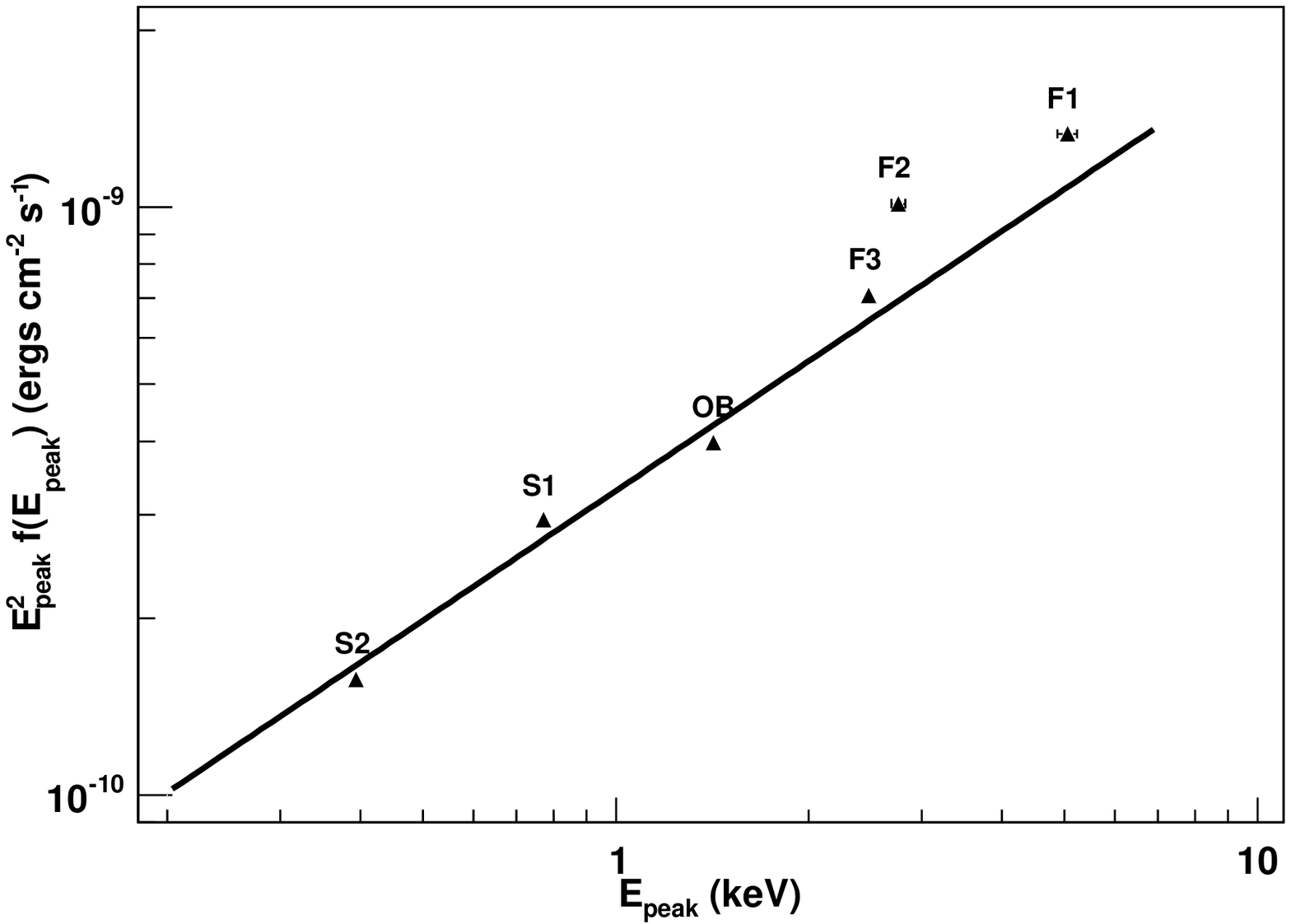}
\caption{
The peak energy and the corresponding flux for different Mrk 421 states,
 as measured by $Swift$-XRT at 0.3$-$10 keV.  The solid line is fitting result using a power-law function, which  yields  $f(E_{peak})=(0.2056\pm0.004) \cdot E_{peak}^{-1.266\pm 0.004}$  keV$^{-1}$ cm$^{-2}$ s$^{-1}$ with $\chi^{2}/dof$=2113/4.
}
\label{fig6}
\vspace*{0.5cm}
\end{figure}

\subsubsection{$MAXI$-GSC and $Swift$-BAT SEDs}
According to $MAXI$-GSC measurements at 2$-$20 keV, the flux
during the flaring periods increases by about a factor 4$-$20, compared to S2.
The spectral indexes for most of
the states are in the 2$-$2.4 range while the spectrum softens (index 2.97$\pm$0.18) during F7.
 In the X-ray band (14$-$195 keV)  the $Swift$-BAT data shows an even larger variation (4$-$70
times) with spectral indexes ranging from 2.5 to 3.1.

\subsubsection{$Fermi$-LAT and ARGO-YBJ SEDs}

As stated in Section 2, we also test the significance of the spectrum
curvature in the $Fermi$-LAT data.
During the S2 phase, the TS$_{curve}$ value is found to be 32.6, corresponding to 5.7 standard deviations (s.d.).
An evidence for a curvature is then observed in S2, with a peak energy E$_{peak}$=(60$\pm$11) GeV.
The TS$_{curve}$ values for F2 and F7 are 8.0 and 9.4, respectively,
corresponding to 2.8 and 3.1 s.d.,
only showing a hint of curvature. The TS$_{curve}$ values for the other seven states are less than 3.4.
No curvature are detected in these cases.
These features are visible in Figure 7. Note that the data points of $Fermi$-LAT
are the result of the analysis made in differential energy ranges, and are independent
of the assumed spectral models.

In the GeV $\gamma$-ray band, the F3, F4, F5, OB, S1 and S2 phases
have similar spectral indexes (ranging from 1.74 to 1.80) and fluxes (within 32\%), as shown in Table 3.
Compared to the S2 state, the spectral index of F1 shows a moderate hardening  (
$\Delta \alpha=$0.20$\pm$0.15) and a flux decrease of  (34$\pm$21)\%.  The spectral index of F2 hardens more significantly ($\Delta \alpha=$0.27$\pm$0.08) with a flux increase of  a factor 2.06$\pm$0.36.
A flux enhancement by a factor 3 is observed during F6 and F7, with a harder
spectral index during F6 ($\Delta \alpha=$0.09$\pm$0.04) and a negligible index variation during F7.

In VHE $\gamma$-ray band the S2 flux is estimated to be (0.33$\pm$0.10) I$_{crab}$,
assuming a fixed spectral index, $\alpha=$2.75.
This result is comparable to the baseline flux of Mrk 421 obtained using a 20 years
long-term combined ACT data \citep{tlucz10}, which is estimated to be less than 0.33 I$_{crab}$
 above 1 TeV.
The averaged measured flux is (0.56$\pm$0.13) and (0.91$\pm$0.14) I$_{crab}$  during S1 and OB phase,
respectively. F2 is the largest flare, achieving a flux of (7.2$\pm$1.5) I$_{crab}$.
The flux of the remaining flares is around (1$-$3) I$_{crab}$.
The spectral index of F7 ($\alpha=$3.22$\pm$0.24) marks the softest spectrum of the observed flares.
The flux modulations appears in coincidence with the X-ray observations.

Summarizing the above results, we can conclude that the flux enhancements are detected in both X-rays
and VHE $\gamma$-rays during all the nine states, compared to the baseline S2. The behavior in the GeV
band is actually different. Accordingly, a phenomenological classification of 3 types of SEDs (T1,
T2 and T3) is here introduced, i.e.,
(I) flares with no or little GeV flux and photon index variations, (II) flares with $\gamma$-ray spectral hardening, irrespective of the flux variations, and (III) flares with flux enhancements, irrespective of spectral behavior.
Type T1 includes phases S1, S2, F3, F4, F5 and OB.
Type T2 includes the F1 and F2 states and also the
day (MJD=56124), corresponding to the F6 maximum flux, during which the spectral index
significantly hardens to $\alpha=$1.60$\pm$0.04.
It is worth to note that this variation is fast: hardening phase only lasts two days and recovers soon,
indicating an unstable state.
Type T3 includes phases F6 and F7. Actually the spectral index of F7 becomes softer above the peak energy
for both low and high energy component. The previous flare on May 7th, 2008, reported by
\cite{accia09b}, not included in this present discussion, may also belong to this type.

During flares of types T1 and T2, the peak energies of both the low and high energy components shift to higher energy with respect to the baseline state S2.  This tendency is consistent with most of the previous measurements \citep{massa08,albert07} based on fragmented observations. This indicates that the modulation of Mrk 421 flux
follows these types in most of the cases.  During flares of type T3, the peak energies could shift to lower energy with respect to S2, but this must be determined by future observations of similar flares.

\begin{figure*}
\includegraphics[width=7in,height=9.in]{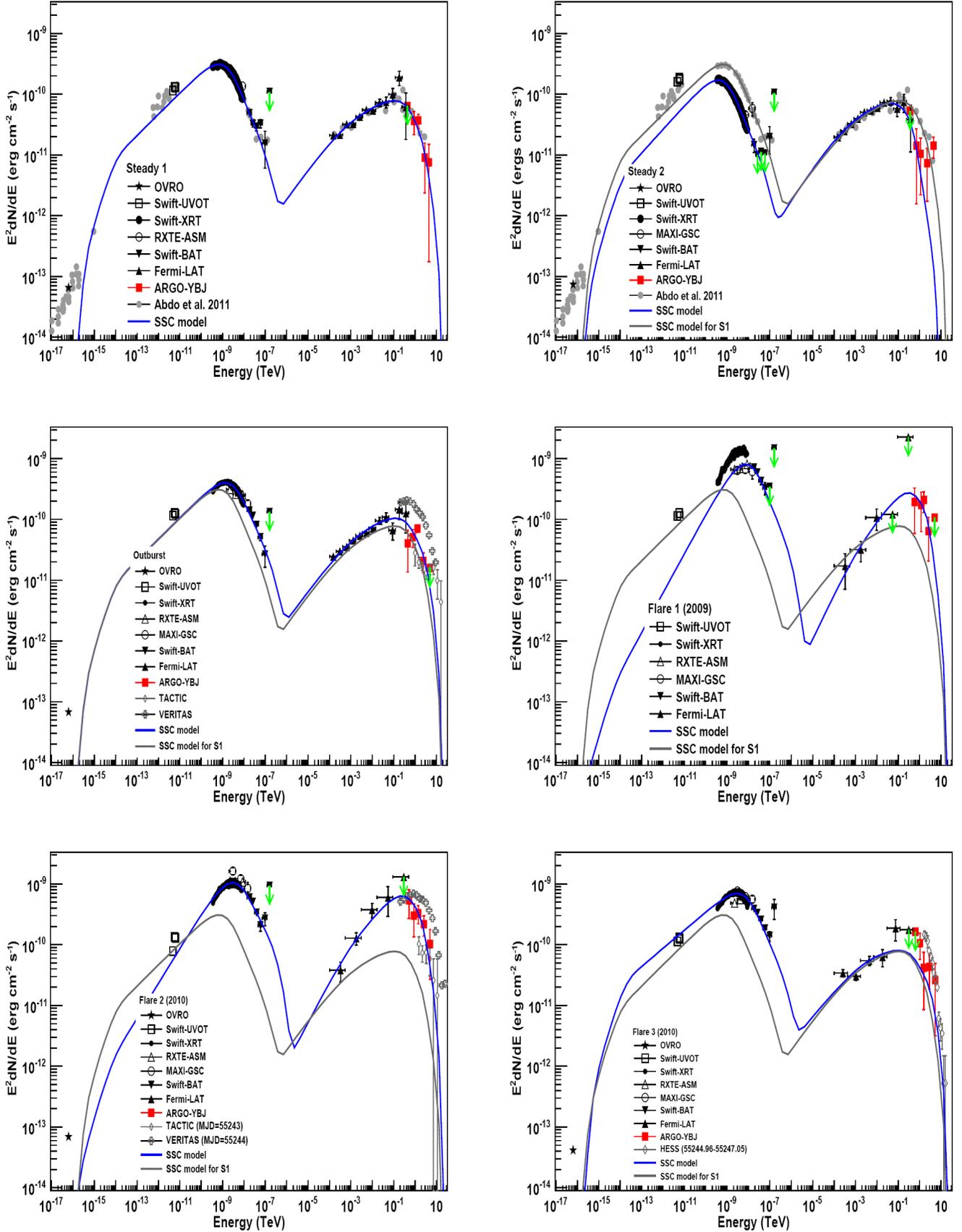}
\caption{
Spectral energy distribution of Mrk 421 during 10 states. The solid line shows the best fit to the data, assuming a homogeneous one-zone SSC model (the best-fit parameters are listed in Table 6).
For comparison, the model describing the   Steady 1 (S1) is also plotted in the other
9 states.
}
\label{fig7}
\vspace*{0.5cm}
\end{figure*}

\begin{figure*}
\includegraphics[width=7in,height=5.5in]{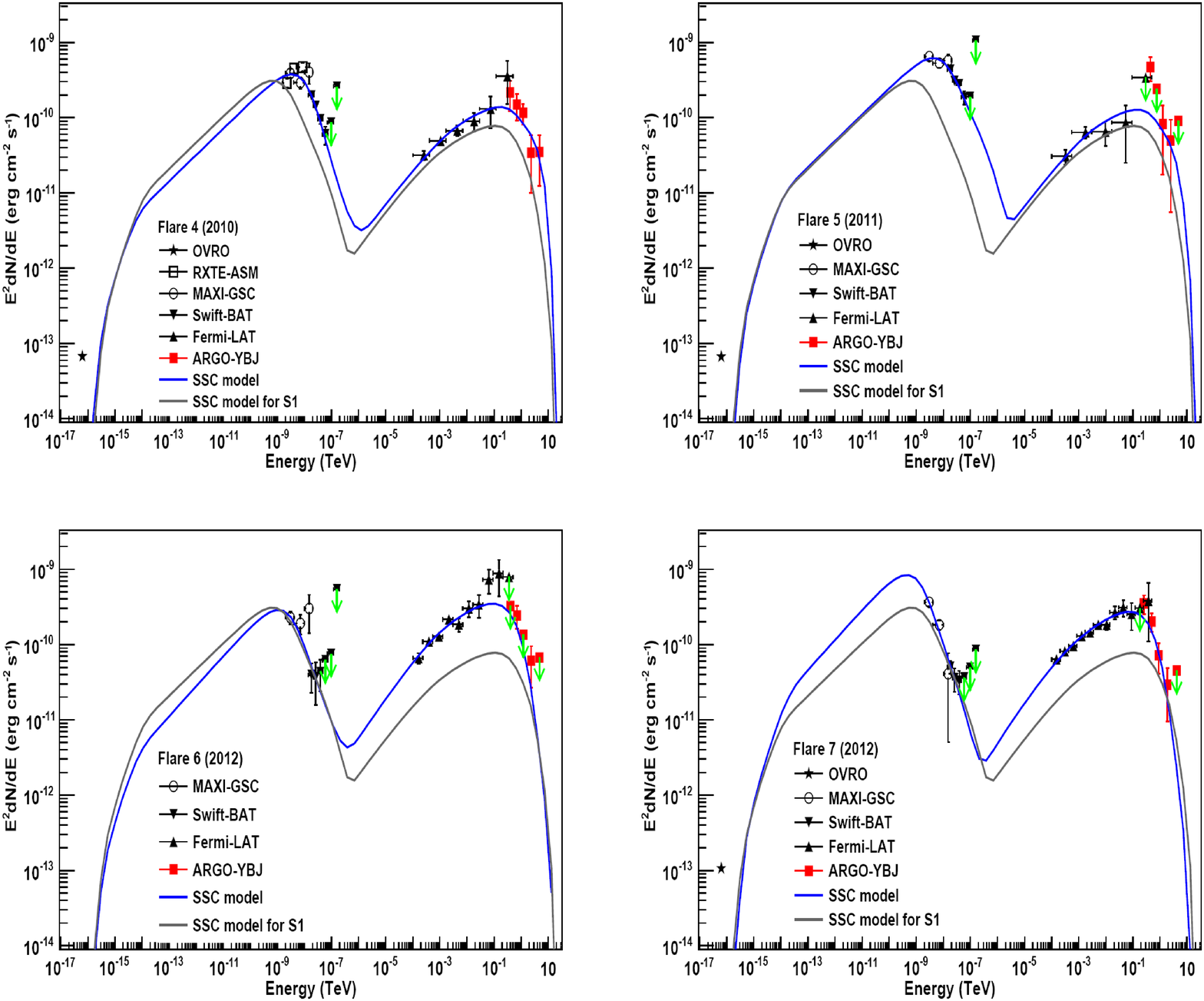}
\caption{Continuation of Figure 7.
}
\label{fig8}
\vspace*{0.5cm}
\end{figure*}

\begin{table*}
\centering
\caption{Energy and  flux of the spectral points
shown in Figure 7 and Figure 8}
\label{tab-1}
\vspace{2mm}
\begin{tabular}{cccccc}
\hline
  State & E       & E$^{2}$dN/dE                & $\bigtriangleup$(E$^{2}$dN/dE)        & 95\% u.l.  & Detector \\
       & (TeV)    &  (erg cm$^{-2}$s$^{-1}$ )  & (erg cm$^{-2}$s$^{-1}$ )  &(erg cm$^{-2}$s$^{-1}$ ) &   \\
\hline
S1 & 4.470e-01   & 1.698e-11 & 2.625e-11 & 6.386e-11 & ARGO-YBJ \\
S1 & 8.910e-01   & 3.632e-11 & 1.494e-11 & 6.094e-11 & ARGO-YBJ \\
S1 & 1.413e+00   & 3.713e-11 & 9.739e-12 & 0 & ARGO-YBJ \\
S1 & 2.818e+00   & 9.018e-12 & 6.650e-12 & 2.024e-11 & ARGO-YBJ \\
S1 & 4.467e+00   & 7.560e-12 & 7.384e-12 & 2.028e-11 & ARGO-YBJ \\
S1 & 1.520e-04   & 2.099e-11 & 1.840e-12 & 0 & Fermi-LAT \\
S1 & 3.080e-04   & 2.143e-11 & 1.447e-12 & 0 & Fermi-LAT \\
S1 & 6.270e-04   & 3.193e-11 & 1.669e-12 & 0 & Fermi-LAT \\
S1 & 1.280e-03   & 3.228e-11 & 1.963e-12 & 0 & Fermi-LAT \\
S1 & 2.590e-03   & 4.266e-11 & 2.837e-12 & 0 & Fermi-LAT \\
S1 & 5.270e-03   & 5.517e-11 & 4.431e-12 & 0 & Fermi-LAT \\
S1 & 1.070e-02   & 5.411e-11 & 6.199e-12 & 0 & Fermi-LAT \\
S1 & 2.180e-02   & 7.172e-11 & 1.020e-11 & 0 & Fermi-LAT \\
S1 & 4.440e-02   & 7.453e-11 & 1.478e-11 & 0 & Fermi-LAT \\
S1 & 9.020e-02   & 9.906e-11 & 2.503e-11 & 0 & Fermi-LAT \\
S1 & 1.830e-01   & 1.883e-10 & 5.048e-11 & 0 & Fermi-LAT \\
S1 & 3.730e-01   & 6.174e-11 & 4.369e-11 & 1.640e-10 & Fermi-LAT \\
S1 & 1.800e-08   & 4.853e-11 & 3.441e-12 & 0 & Swift-BAT \\
S1 & 2.605e-08   & 3.283e-11 & 4.348e-12 & 0 & Swift-BAT \\
S1 & 3.845e-08   & 2.889e-11 & 4.074e-12 & 0 & Swift-BAT \\
\hline
\multicolumn{6}{l}{This table is available in its entirety in a machine-readable form  in the online journal. A portion is }\\
\multicolumn{6}{l}{ shown here for guidance regarding its form and content.}
\end{tabular}
\vspace*{0.5cm}
\end{table*}

\subsection{Cherenkov detectors VHE  $\gamma$-ray data}

\begin{figure*}
\includegraphics[width=7in,height=2.5in]{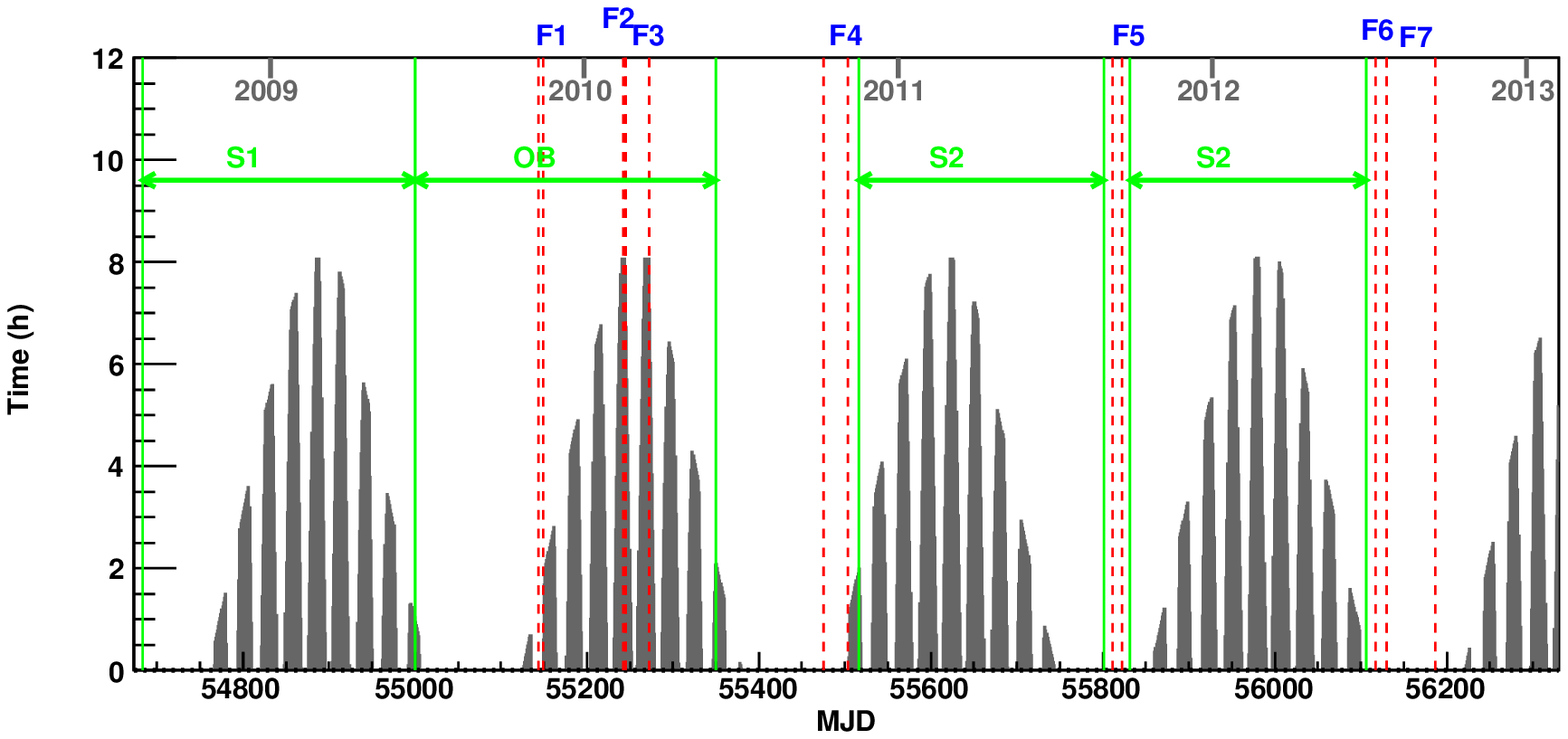}
\caption{
The daily allowable Mrk 421 observation time for a Cherenkov detector located at 30$^{\circ}$ North of latitude, during the 4.5 years considered in this work.
}
\label{fig9}
\vspace*{0.5cm}
\end{figure*}
During the 4.5 years period from August 2008 to February 2013, Cherenkov detectors (e.g. VERITAS, MAGIC),
also observed Mrk 421 in the VHE band. Even if these detectors cannot monitor Mrk 421 day by day,
they can provide more precise measurements for short periods, due to their excellent sensitivity.

To give an example of the timing coverage of a Cherenkov detector, a dummy instrument
located at 30$^{\circ}$  N (close to the latitude of VERITAS, 32$^{\circ}$ N, and MAGIC, 28$^{\circ}$  N) is here considered to estimate the allowable observation time for Mrk 421.
Figure 9 shows the allowable time, requiring
the Sun zenith angle be greater than 105$^{\circ}$, the Moon zenith angle greater than 100$^{\circ}$, and the Mrk 421 zenith angle less than 50$^{\circ}$.

Among the 7 flares presented in the last section, F4, F5, F6 and F7 occurred during the period from July to October, forbidden for Cherenkov detector observations, since Mrk 421 is close to the direction of the Sun. The moonlight completely hampered Cherenkov telescope observations during F1 and
partially during F3. Only flare F2 could be eventually observed every night by Cherenkov detectors.

Actually VERITAS observed Mrk 421 during the last day of F2, i.e. MJD=55244.
The preliminary spectrum is shown in Figure 7 and the corresponding flux above 1 TeV is 16.8 I$_{crab}$ \citep{fort12}.
TACTIC  observed Mrk 421 every night during F2 \citep{singh15}. The peak flux measured on February 16 (MJD=55243) and the corresponding spectrum during the peak flux day is also shown in Figure 7. The flux above 1 TeV is 2.7 I$_{crab}$. HAGAR also observed Mrk 421 every night during F2 \citep{shukla12}. The peak flux on February 17 (MJD=55244) is about 7 I$_{crab}$. It is worth to recall that the average flux detected by ARGO-YBJ during the three days of F2 is (7.2$\pm$1.5) I$_{crab}$.

Both VERITAS and HESS observed the first three days of flare F3. VERITAS preliminary results show that the flux decreased from 5.7 I$_{crab}$ to 2.9  I$_{crab}$ \citep{fort12}. HESS preliminary results also shows a decreasing flux from 4.8 to 1.4 I$_{crab}$ \citep{tlucz11}.

The S1 period extends over years 2008$-$2009 and during this time the
Whipple telescope monitored the Mrk 421 emission.
The total observation time is 130.6 hours and the mean $\gamma$-ray rate is 0.55$\pm$0.03 I$_{crab}$ \citep{accia14}, which is consistent with the ARGO-YBJ result, i.e., 0.56$\pm$0.13 I$_{crab}$.
During this period, MAGIC also observed Mrk 421 for about 27.7 hours \citep{abdo11,aleksi15}.
The  photon fluxes for the individual observations gave an average flux of about 50\% that
of the Crab Nebula, with relatively mild (typically less than a factor of two) flux variations.
The spectrum,   shown in Figure 7,  is consistent with the ARGO-YBJ data.

The OB period lasts over the years 2009$-$2010. The spectrum measured by
TACTIC, shown in Figure 7, is lower than that by ARGO-YBJ. The
preliminary spectrum reported by VERITAS \citep{galan11} is also
shown in Figure 7, which is higher than the one of ARGO-YBJ. These
differences may be caused by the different observation times, considering that
Mrk 421 was in an active and variable phase.

\section{The SSC model for Mrk 421}
The different types of Mrk 421  flux variations can be associated to the intrinsic astrophysical
mechanisms of the emission. In the following, we will investigate the major parameters correlated
to these variations in the framework of the one-zone SSC emission model.

%\subsection{Fitting the SED}
In this paper the one-zone homogeneous SSC model proposed by \cite{krawczy04} is
adopted to fit the multi-wavelength SEDs measured during different states. In this model a spherical
blob of plasma with a co-moving radius $R$ is assumed. The relativistic Doppler Factor of the
emitting plasma is defined as  $\delta =[\Gamma(1-\beta cos\theta)]^{-1}$, where $\Gamma$ is the bulk Lorentz factor of the
emitting plasma, $\beta$ is its bulk velocity in units of the speed of light, and $\theta$ is the angle between the jet axis and the line of sight, as measured in the observer frame.
The emission volume is filled with an isotropic population of electrons and a randomly oriented
uniform magnetic field $B$. The SED of the injected electrons in the jet frame is assumed to follow a
broken power law with indexes $p1$ and $p2$ below and above the break energy $E_{break}$.
The electron distribution is normalized by a factor $u_{e}$ (in units of erg cm$^{-3}$).
To reduce the free parameters in the model, the low limit for the electron energy $E_{min}$ is fixed to be 500$\cdot m_{e}c^{2}$
 and the high limit $E_{max}$ is assumed to be 10$\cdot E_{break}$.
The radius of the emission zone is constrained by the variability of the time scale as $R<ct_{var}\delta/(1+z)$.
In the multi-wavelength data considered in this analysis, the fastest variability has a time scale of   $\sim$1 day,  observed during F2, in X-rays and GeV $\gamma$-rays, as shown in Figure 4. In this work, the radius $R$ for
all phases are arbitrarily set to be 10$^{14}$ m,  being $t_{var}>4.8(20/\delta)$ hours the allowed time
variability range.
We have so far still six free parameters: $p1$, $p2$, E$_{break}$,  $u_{e}$,  B and $\delta$
to be determined experimentally by fitting the data presented in Figure 7.

In this work, we use the least-square method to  determine the best values of the parameters $p1$, $p2$, E$_{br}$, B, $\delta$, and $u_{e}$. For the $Swift$-XRT flux data, we   added a 3\% systematic error besides the statistical  error listed in Table 5.  The ultraviolet and radio
data points were not used for the fit, as will be discussed later.
The Extragalactic Background Light (EBL)
absorption of the VHE $\gamma$-ray is included in the calculation, according to the \cite{franc08} model.

Since the parameter $p1$ is related to the spectral shape at energies below the synchrotron and inverse Compton peaks, it is mainly determined by $Swift$-XRT data in X-rays and by $Fermi$-LAT data in  GeV $\gamma$-rays.
For five out of ten states the $Swift$-XRT data around the synchrotron peak are not available, therefore in this cases  $p1$ is mainly determined by the $Fermi$-LAT data.  The synchrotron peak energies of states S1 and S2  are close to the low energy limit of the $Swift$-XRT data, hence  the measurement below the synchrotron peak cannot strongly constrain the $p1$ value, that also in this case will be mainly determined by the $Fermi$-LAT data.
For the remaining three states, the statistical accuracy  of the $Swift$-XRT data is much better than that of other detectors,   however  the low energy $Swift$-XRT measurements (at energies below $\sim$1 keV) are      biased by the uncertainty in the absorption of hydrogen-equivalent column density and  by   detector systematics  \footnote{http://heasarc.gsfc.nasa.gov/docs/heasarc/caldb/swift/docs/xrt/}(such as the CCD charge trapping, generated by radiation and high-energy proton damage, affecting mostly the low energy events).
For this reason we prefer to use for all the 10 states only $\gamma$-ray data when fitting the parameter $p1$. The result and  chi squares ($\chi^{2}_{\gamma}$/ndf) obtained by the fitting procedure are listed in Table 6. The derived $p1$ values for type T1 and T3 states  are consistent within errors,   ranging from 2.2 to 2.4. The results for T2 flares indicate a harder electron spectrum,   with $p1=1.7\pm0.3$ for F1 and $p1=1.85\pm0.20$ for F2.

To obtain the remaining 5 parameters, the whole SED data above 0.3 keV are used.  The $Swift$-XRT observation
during F1 is not taken into account.
 The parameter $p2$  is determined by the spectral shape above  the synchrotron and inverse Compton peak energies. Therefore,
the parameter $p2$ is mainly determined by the  $Swift$-XRT, $RXTE$-ASM, $MAXI$-GSC and $Swift$-BAT data in X-rays, and by the ARGO-YBJ data in TeV $\gamma$-rays.
The accuracy of   $Swift$-XRT measurements is much higher than that of other detectors, therefore   $p2$ can be well determined for states in which $Swift$-XRT data are available.
The parameter  $\delta$ is mainly determined by both  the synchrotron and inverse Compton peak energies.
There are no data   data around the synchrotron peak for F5, F6 and F7, therefore in these cases $\delta$ cannot be well constrained. In particular, we cannot find a best value for F7 within a reasonable range. For flare F7, we arbitrarily set the value of $E_{break}$ equal to the one derived for the S2 state.

According to the fit results, the SSC model reasonably describes all the SEDs, as shown in Figure 7 and 8.
The fact that $Swift$-XRT data below 0.5 keV are not perfectly
described during OB, F2 and F3, may be explained by the systematic errors previously discussed.
The obtained parameters  and chi squares ($\chi^{2}_{all}$/ndf) are summarized in Table 6.
The  Doppler Factor $\delta$, ranging from 10 to 41, is similar to the ones found in previous
investigations \citep{abdo11,barto11,shukla12,zhang12}. The
values of the magnetic field, $B$ $\sim$0.1 G, are almost constant within a factor 2. The jet power in electrons ($10^{46}-10^{47}$ erg) is much higher than that in magnetic field, as indicated by the electron energy density to the magnetic field energy density ratio $u_{e}/u_{B}$, listed in Table 6.

\begin{table*}
\centering
\caption{Best-Fit parameters in the SSC Model  for 10 states}
\label{tab-1}
\vspace{2mm}
\begin{tabular}{c|ccc|cc|ccc|cc}
\hline
  State& $p_1$  & $p_2$    & log(E$_{break}$) & $\delta$ & B  & $u_{e}$   & $u_{e}/u_{B}$ $^{a}$ &W$_{e}$ $^{b}$ &$\chi^{2}_{\gamma}$/ndf & $\chi^{2}_{all}$/ndf \\
                &        &          &  (eV)            &          &(G) &($10^{3}$erg cm$^{-3}$) & &($10^{46}$erg) & &\\
\hline
Steady 1&  2.30$_{-0.04}^{+0.06}$  &4.70$_{-0.03}^{+0.07}$  & 11.00$_{-0.03}^{+0.04}$   &38$_{-4}^{+6}$   &0.048$_{-0.012}^{+0.012}$   &6.65$_{-0.15}^{+0.15}$    &70.6 &2.8 &23.0/11 &238/501\\
\hline
Steady 2&  2.22$_{-0.06}^{+0.09}$  &4.68$_{-0.04}^{+0.02}$ & 10.78$_{-0.04}^{+0.05}$    &15$_{-2}^{+4}$   &0.17$_{-0.05}^{+0.07}$   &12.7$_{-0.5}^{+0.4}$   &10.6 &5.3 &13.2/11 &236/467\\
\hline
Outburst&  2.30$_{-0.05}^{+0.08}$ &4.51$_{-0.02}^{+0.03}$ & 11.13$_{-0.05}^{+0.02}$    &35$_{-5}^{+3}$   &0.054$_{-0.005}^{+0.026}$   &7.74$_{-0.14}^{+0.24}$    &65.7  &3.2& 23.5/12&600.8/643\\
\hline
Flare 1 &  1.7$_{-0.3}^{+0.3}$  &4.7$_{-0.5}^{+1.2}$ & 11.51$_{-0.03}^{+0.09}$    &10$_{-2}^{+2}$   &0.14$_{-0.04}^{+0.07}$   &31$_{-5}^{+7}$    &41.3  &12.9& 1.6/2 & 7.3/14\\
\hline
Flare 2 &  1.85$_{-0.20}^{+0.20}$  &4.30$_{-0.12}^{+0.05}$  & 11.27$_{-0.03}^{+0.03}$    &17$_{-2}^{+3}$   &0.092$_{-0.024}^{+0.028}$    &24$_{-3}^{+2}$   &73.0 &10.1& 3.7/3& 179/308\\
\hline
Flare 3 &  2.40$_{-0.15}^{+0.15}$  &4.60$_{-0.09}^{+0.08}$ & 11.21$_{-0.03}^{+0.02}$   &41$_{-3}^{+5}$   &0.080$_{-0.017}^{+0.011}$  &4.40$_{-0.10}^{+0.07}$    &17.3 &1.8 &13.2/4 & 1055.7/574\\
\hline
Flare 4 &  2.30$_{-0.15}^{+0.15}$  &5.6$_{-0.6}^{+0.9}$ & 11.49$_{-0.07}^{+0.08}$   &35$_{-7}^{+10}$   &0.033$_{-0.013}^{+0.019}$   &12.3$_{-1.0}^{+1.3}$   &289.2&5.2 & 6.1/5 &23.4/15\\
\hline
Flare 5 &  2.3$_{-0.2}^{+0.4}$  &4.6$_{-0.5}^{+0.7}$ & 11.35$_{-0.12}^{+0.13}$   &31$_{-13}^{+21}$   &0.072$_{-0.047}^{+0.108}$   &8.0$_{-1.2}^{+1.4}$    &39.2 &3.4&6.8/4 &12.1/13\\
\hline
Flare 6 &  2.20$_{-0.17}^{+0.11}$  &5.1$_{-0.7}^{+1.8}$ & 11.17$_{-0.43}^{+0.24}$   &15$_{-5}^{+24}$   &0.085$_{-0.033}^{+0.053}$   &40$_{-30}^{+28}$   &120.8  &16.8& 12.1/8 & 17.5/17\\
\hline
Flare 7 &  2.20$_{-0.07}^{+0.12}$  &5.2$_{-0.2}^{+0.3}$ & 10.78     &30$_{-5}^{+7}$   &0.115$_{-0.032}^{+0.038}$   &8.77$_{-1.32}^{+1.40}$   &16.6 &3.7 &2.45/3  & 14.1/20\\
\hline
\multicolumn{9}{l}{$^{a}$ $u_{e}/u_{B}$=8$\pi$$u_{e}$/B$^{2}$}\\
\multicolumn{9}{l}{$^{b}$ $W_{e}=u_{e} \frac{4}{3}\pi R^{3}$}\\
\end{tabular}
\vspace*{0.5cm}
\end{table*}

\section{Discussion}
Using the long-term multi-wavelength data from radio to VHE $\gamma$-rays, we have shown that Mrk 421 is active at all wavebands during the 4.5 years considered in this work. The variability of Mrk 421 increases with energy for both the low and high energy SED components.
The source is highly variable in the X-ray and VHE $\gamma$-ray bands, with the normalized variability amplitudes greater than 70\% (see Figure 3).
An overall cross-correlation analysis (see Table 2)  between $Swift$-BAT   and ARGO-YBJ data shows that the variabilities in X-rays and VHE $\gamma$-rays  are generally correlated. The  correlation is also clearly visible in the light curves during the large X-ray flares (see Figure 2, 4 and 5).
Our previous observations during the outburst of 2008 also show that   X-rays and VHE $\gamma$-rays were tightly correlated with the peak times  in good agreement with each other \citep{barto11}. A clear correlation between    X-rays and VHE $\gamma$-rays  has also been reported in many observations in the past decades \citep{blaze05,chen13,accia14}. All these results firmly support the idea that the X-ray and VHE $\gamma$-ray emissions of Mrk 421 originate from the same zone.

 The variability amplitude in   GeV $\gamma$-rays is   39\% (see Figure 3), which is less than that in VHE $\gamma$-rays.
In fact, the amplitude of the GeV $\gamma$-ray variability  is very low during most of the time, being only about 20\% if the large GeV $\gamma$-ray flares  F6 and F7 are excluded.
The overall cross-correlation analysis (see Table 2) shows that   GeV $\gamma$-rays are  moderately correlated with VHE $\gamma$-rays.
The GeV and VHE $\gamma$-ray light curves reported in Figure 4 and 5 show an evident correlation only during F2, F6 and F7. According to the measurements shown in Table 3, during flares F3, F4, F5 and OB phase, the flux above 0.1 GeV increases by 20\%-46\% with respect to the S1 steady state, while the VHE flux increases by 63\%-275\%. With these results, we could not therefore exclude the possibility that GeV and TeV $\gamma$-rays are produced in different emission zones. A possible scenario would
 be that the observed emission is the superposition of one stationary
zone with a steady GeV emission and one active zone responsible for the VHE flux variation  (e.g. \cite{cao13,aleksi15b}).
 However, during flares F1, F2, F6, and F7, GeV and TeV $\gamma$-rays   show clear correlated
variations, according to the SEDs reported in Figure 7.
This could indicate that GeV and TeV emissions are generated in the same zone,
at least during these phases. In particular during flare F1
the flux above 0.1 GeV  decreases of about (27$\pm$24)\%.
If this decrease is  not due to a statistical fluctuation, the stationary zone would be excluded, and the above hypothesis of two emission zones could be excluded too. Moreover,
during F6 and F7, the GeV $\gamma$-ray flux increases by  a large amount  (240\%$-$295\%), while  the spectral indices were about consistent.
Also this result   does not support the hypothesis of two zones, since the predicted spectrum at GeV energies for the active zone is much harder than that for the steady zone, according to \cite{cao13} and \cite{aleksi15b}.

Compared to other wavebands, the flux variation at radio frequencies is much weaker. A radio flare is observed in 2012, which is the largest radio flare ever observed in Mrk 421 \citep{hovatta15}.  The cross-correlation analysis shows that this radio flare
follows the GeV $\gamma$-ray flare with a time-lag of about 42 days (see Table 2), which is consistent with \cite{hovatta15}. If the radio and GeV $\gamma$-ray flares are physically connected \citep{hovatta15}, then the GeV $\gamma$-ray emission could originate upstream of the radio emission. The distance between radio and GeV $\gamma$-ray emission sites and their distances to the central black hole have been discussed by \cite{max14}.

The variability amplitude in the UV band is 33\%, similar to that in GeV $\gamma$-rays, however according to the cross-correlation analysis (see Table 2) and to the light curves (see Figure 2 and 4) the UV flux appears not to be correlated with GeV $\gamma$-rays. The UV flux does not even shown an evident  correlation with X-rays. Moreover, according to the SED of flare F1 shown in Figure 7, the UV and X-ray data cannot be fitted together with a unique component. These results seem to indicate that the UV and X-ray emissions do not share the same origin. Instead, the moderate correlation observed between UV and radio suggests that their emission regions could be the same, possibly located downstream of the X-ray and $\gamma$-ray emission zone.
Therefore, the  radio and UV
data points were removed from the SSC model fitting. This choice is different from other authors, e.g. \citep{abdo11,shukla12} who included them when fitting the SED. Finally, the SSC model generally under-predicts the UV and radio emissions.
This supports our  hypothesis that,  at least  partially, the radio and UV
emission comes from regions in the jet not emitting X-rays and $\gamma$-rays.

For the type T1 and T3 states, the derived spectral indices $p1$ are generally consistent with $p1$=2.2, i.e. the canonical particle spectral index
predicted for relativistic diffuse shock acceleration (for a review see \cite{kirk99}). This
suggests that this process is active in Mrk 421. The change of the electron index
$\Delta p$ = $p2$ - $p1$  is larger than the expected typical cooling break
$\Delta p$ =1 \citep{kino02}, indicating that the break is not induced by radiative cooling. The cooling mechanism may be less important, as suggested in \cite{accia11}. \cite{abdo11} speculated that the steep break is a characteristic of the acceleration process which is not yet
understood. The break energy $E_{break}$ should represent the maximum energy that can be achieved
in the acceleration process, depending on the acceleration time  of the particle in the shock area.
Accordingly,  the flares of  T1 type  should be mainly caused by the variation of the maximum energy of
the electrons reached within the shock area.  Concerning the type T3 flares (i.e., F6 and F7), these flares might be   due to the  increase of  the magnetic field (B) or comoving particle density ($u_e$)    compared to S1.

Up to now, it is not yet clear how shocks work in a jet. If the different states detected in this work are caused by different shocks, we could expect different features of emission zones, such as B, $\delta$, R, electron density and spectrum. If the different states are caused by similar shocks moving down the jet, we would expect emission zones with the same  R, for all the states, as we assumed when fitting the SED using the SSC model. In such a hypothesis, we could guess that the underlying mechanism responsible for the flares of T1 and T3 types may be due to the variation of the ambient medium. In the theory of diffusive acceleration, the acceleration time scale of particles is related to the strength of the magnetic field both in upstream and downstream regions   \citep{drury83}. The upstream magnetic field could be related to the ambient medium. The variation of the acceleration time scale may change the maximum energy of the particles achieved within the shock. When the density and the magnetic field of the ambient medium are different, the number of particles and the corresponding maximum energy can be different. In such a hypothesis, the variability time scale for each state would characterize the size of each medium block that is crossed by the shock.

The T2 flares require a harder injected electron spectrum than the other types. This change would be caused by the acceleration processes. The slopes of F1 and F2 go beyond the predictions of a spectral index of 2.0 given by the canonical non-relativistic diffuse shock acceleration. According to \cite{steck07}, particle spectra with spectral indexes less than 2 can be realized within the relativistic shock using extreme parameters, i.e., large scattering angles. F1 and F2 flares only last a few days, which means that the produced electrons with spectral index less than 2.2 is a transient state and cannot last for a long time. So, we cannot exclude the possibility that such short flares are due to extreme parameters. Another alternative mechanism for such a hard spectrum would be that the particles accelerated by the shock are subsequently accelerated by the stochastic process in the downstream region, which is able to produce spectra that are significantly harder than the limits of the first-order mechanism within a short time \citep{virtan05}. Recently, \cite{guo14} also predicts a hard spectral index resulting from relativistic magnetic reconnection.

\section{Summary}
In this paper, we have presented a 4.5-year    continuous   multi-wavelength monitoring of Mrk 421, from 2008 August 5  to  2013 February 7, a period that includes   both   steady states and episodes of strong flaring activity. The observations concern a wide energy range, from  radio to TeV $\gamma$-rays.
In particular, thanks to the ARGO-YBJ and $Fermi$ data, the whole energy range from 100 MeV to 10 TeV is covered without any gap.
These extensive datasets are essential for
studying   the origin of the flux variability and to investigate the underlying emission mechanism.   The main results of this work can be summarized
as follows:

1. Mrk 421 showed both low and high activity phases at all wavebands during the 4.5 year period analyzed in this work. The variability increases with energy for both the SED components. Concerning the synchrotron component, the variability amplitude increases from 21\% in radio and 33\% in UV to 71\%-73\% in soft X-rays and 103\%-137\% in hard X-rays. For the Inverse Compton component, the amplitude is 39\% at GeV energies and increases to 84\% at TeV energies.

2. The time correlation among the flux variation in different wavebands was analyzed.  The variation of the X-ray flux is clearly correlated with the TeV $\gamma$-ray flux.
This result is consistent with many previous observations in the past decades, supporting the idea  that the X-ray and VHE $\gamma$-ray emissions   originate from the same zone.  The GeV $\gamma$-ray flux appears to be  moderately correlated with the TeV $\gamma$-ray flux. This result is new  compared to previous results. The correlation is mainly due to the large GeV $\gamma$-ray flares occurred in 2012 and a large X-ray/TeV $\gamma$-ray flare in 2010. Taking into account the spectral shape during this flares, we can conclude that the GeV and TeV $\gamma$-rays also originate from the same zone, at least during these flares. On the contrary,  X-ray and $\gamma$-ray fluxes are weakly or not correlated with radio and UV fluxes.

3. According to the observed light curves, ten states (two steady periods, S1 and S2, one outburst period, OB, and seven large flares) have been selected and analyzed. Five flares have been identified in X-rays, and two in GeV $\gamma$-rays. The duration of the flares  ranges between  3 and 58 days.  X-ray and TeV $\gamma$-ray fluxes increase during all  the active states.
In X-rays, the flux increases by a factor  4$-$70 and the peak energy increases from 0.4 keV  to   1.4$-$5.1 keV. At TeV energies, the flux has been observed to vary from 0.33 I$_{crab}$ to 7 I$_{crab}$.

4. According to the behavior of GeV $\gamma$-rays, the activity states can be classified    into three groups.
(I) flares with no or little GeV flux and photon index variations, e.g.  S1, S2, OB, F3, F4 and F5; (II) flares with $\gamma$-ray spectral hardening, irrespective of the flux variations, e.g. F1 and F2;  (III) flares with flux enhancements, irrespective of spectral behavior, e.g. F6 and F7.

5. A simple one-zone SSC model is adopted to fit the multi-wavelength SED state by state. The SSC model can satisfactorily   reproduce all the  SED measurement except the $Swift$-XRT data below 0.5 keV, probably due to a detector systematics affecting the low energy events. For type I and III states, the derived injected electron spectral indices are around 2.2, as expected from relativistic diffuse shock acceleration, indicating that this process can be active in Mrk 421. According to the derived parameters, the variation of these states may be caused by the variation of   environment properties. Instead,  type II flares  require  harder injected electron spectra, with spectral indices around 1.7$-$1.9.  The underlying physical mechanisms responsible for this type of flares may be related to the acceleration process itself.

\acknowledgments
 This work is supported in China by NSFC (No.11205165, No.11575203, No.11375210),
the Chinese Ministry of Science and Technology, the
Chinese Academy of Sciences, the Key Laboratory of Particle
Astrophysics, CAS, and in Italy by the Istituto Nazionale di Fisica
Nucleare (INFN).
We are grateful to   Dahai Yan and Liang Chen for their helpful   suggestions that improved the paper.
We also acknowledge the essential supports of W.Y. Chen, G. Yang,
X.F. Yuan, C.Y. Zhao, R. Assiro, B. Biondo, S. Bricola, F. Budano,
A. Corvaglia, B. D'Aquino, R. Esposito, A. Innocente, A. Mangano,
E. Pastori, C. Pinto, E. Reali, F. Taurino and A. Zerbini, in the
installation, debugging and maintenance of the detector of ARGO-YBJ.
This research has made use of data and  software provided by the High Energy Astrophysics Science Archive Research Center (HEASARC), which is a service of the Astrophysics Science Division at NASA/GSFC and the High Energy Astrophysics Division of the Smithsonian Astrophysical Observatory.
We acknowledge the use of  the ASM/RXTE quick-look results provided by the ASM/RXTE team, the MAXI data provided by RIKEN, JAXA and the MAXI team, Swift/BAT transient monitor results provided by the Swift/BAT team, and the  15 GHz  data provided by the OVRO team.

%\vspace{\baselineskip}

\clearpage

\end{document}